\documentclass{article}
\usepackage{kr}

\usepackage{times}
\usepackage{soul}
\usepackage{url}
\usepackage[hidelinks]{hyperref}
\usepackage[utf8]{inputenc}
\usepackage[small]{caption}
\usepackage{graphicx}
\usepackage{amsmath}
\usepackage{amsthm}
\usepackage{booktabs}
\usepackage{algorithm}
\usepackage{algorithmic}
\title{BLAST: Benchmarking LLMs with ASP-based Structured Testing}

\author{%
    Manuel Alejandro Borroto Santana$^1$\thanks{Equal contribution.}\and
    Erica Coppolillo$^{1,2*}$\and
    \\
    Francesco Calimeri$^{1,3,4}$\and 
    Giuseppe Manco$^1$\and 
    Simona Perri$^3$\and 
    Francesco Ricca$^3$
    \affiliations
    $^1$University of Calabria, Rende, Italy\\
    $^2$ICAR-CNR, Rende, Italy\\
    $^3$Artificial Intelligence LAB, DeMaCS, University of Calabria, Rende, Italy\\
    $^4$DLVSystem, Rende, Italy\\
    \emails
     \{name.surname\}@unical.it,
     \{name.surname\}@icar.cnr.it
}

\usepackage{algorithm}
\usepackage{algorithmic}
\usepackage{booktabs}
\usepackage{multirow}
\usepackage[inline]{enumitem}
\usepackage{url}

\usepackage{natbib}
\usepackage[framemethod=tikz]{mdframed}
\usepackage{amssymb}

\usepackage{amsmath}
\usepackage{graphicx}
\usepackage{multirow}

\usepackage{amsmath,amsfonts}
\usepackage{enumitem}
\usepackage[framemethod=tikz]{mdframed}
\usepackage{todonotes}
\usepackage{xspace}

\usepackage{graphicx}
\usepackage[framemethod=tikz]{mdframed}
\usepackage{subcaption}

\usepackage{caption}
\usepackage{float}

\usepackage{xargs}
\usepackage{xcolor,colortbl}
\usepackage{cleveref}

\makeatletter
\newcommand\footnoteref[1]{\protected@xdef\@thefnmark{\ref{#1}}\@footnotemark}
\makeatother

\definecolor{Gray}{gray}{0.85}
\newcolumntype{a}{>{\columncolor{Gray}}c}

\usepackage{algorithm}
\usepackage{algorithmic}
\usepackage{booktabs}
\usepackage{multirow}
\usepackage[inline]{enumitem}
\usepackage{url}
\usepackage{listings}

\newcommand{\llmprob}[2]{\ensuremath{P\left(#1|#2\right)}}

\newcommand{\x}{\ensuremath{x}\xspace}
\newcommand{\y}{\ensuremath{y}\xspace}

\begin{document}

\maketitle

\begin{abstract}

Large Language Models (LLMs) have demonstrated remarkable performance across a broad spectrum of tasks, including natural language understanding, dialogue systems, and code generation. 
Despite evident progress, less attention has been paid to their effectiveness in handling declarative paradigms such as Answer Set Programming (ASP), to date. In this paper we introduce BLAST: The first dedicated benchmarking methodology and associated dataset for evaluating the accuracy of LLMs in generating ASP code. BLAST provides a structured evaluation framework featuring two novel semantic metrics tailored to ASP code generation.
The paper presents the results of an empirical evaluation involving ten well-established graph-related problems from the ASP literature and a diverse set of eight state-of-the-art LLMs.
\end{abstract}


\section{Introduction}\label{sec:intro}
Large Language Models (LLMs)~\citep{zhao2025surveylargelanguagemodels} have been showing remarkable performance across a broad spectrum of tasks, including natural language understanding~\citep{qin2024largelanguagemodelsmeet}, dialogue systems~\citep{ou-etal-2024-dialogbench}, and code generation~\citep{jiang2024surveylargelanguagemodels}. 
More recently, some efforts have been made to encompass advanced reasoning, knowledge representation, and logical formalisms~\citep{ijcai24-borroto,DBLP:conf/esws/LippolisSKZCGBN25,DBLP:journals/tmlr/ValmeekamSGK25,DBLP:journals/corr/abs-2404-07751,coppolilloLLASP}.
In this context, while several benchmarks have been proposed to assess LLM performance on imperative and web-oriented programming languages (e.g., C++, Java, HTML)~\citep{DBLP:journals/software/ErnstBM22,DBLP:journals/corr/abs-2302-06590,DBLP:journals/jss/DakhelMNKDJ23}, far less attention has been given to declarative paradigms such as Answer Set Programming (ASP)~\citep{DBLP:journals/cacm/BrewkaET11,DBLP:journals/ngc/GelfondL91}, despite some recent initial efforts~\citep{ijcai24-borroto,leveraging-llms,coppolilloLLASP,DBLP:journals/corr/abs-2512-17093}.

Rooted in logic programming and nonmonotonic reasoning, ASP is well-suited for modeling and solving complex AI problems~\citep{DBLP:journals/cacm/BrewkaET11}.
Over the years, ASP became increasingly adopted for its expressive language and the availability of efficient and reliable solvers~\citep{DBLP:journals/tplp/GebserMR20}, notably Clingo~\citep{DBLP:conf/iclp/GebserKKOSW16} and DLV~\citep{DBLP:conf/lpnmr/AlvianoCDFLPRVZ17}. 
ASP as fully declarative paradigm has been successfully employed in both academic research and industrial applications, particularly in complex domains (e.g., scheduling, configuration, robotics, workforce planning, decision support)~\citep{DBLP:journals/aim/ErdemGL16,DBLP:journals/ki/FalknerFSTT18}.
In recent years substantial effort has been devoted to developing programming environments and tools that support the creation of ASP specifications, including advanced editors, debuggers, testing frameworks, and utilities~\citep{
DBLP:conf/iclp/AlvianoCR23,DBLP:journals/corr/abs-2303-10118,DBLP:conf/lpnmr/FebbraroRR11,DBLP:journals/tplp/BusoniuOPST13,DBLP:journals/ki/CalimeriGPRR18}. 

\noindent\textbf{{Motivation.}} Despite these advancements, writing ASP code remains often challenging, especially for newcomers without a solid background in logic and mathematics. 
%
In this context, LLMs represent a promising avenue for supporting, and even automating, ASP coding. 
Recent studies have highlighted significant limitations in the ability of general-purpose LLMs to generate correct and semantically sound ASP programs~\citep{coppolilloLLASP}. At the same time, other lines of research, including approaches specifically tailored to ASP encoding demonstrate that the overall direction is viable and promising~\citep{leveraging-llms,ijcai24-borroto,coppolilloLLASP,DBLP:journals/corr/abs-2512-17093}.
However, the field still lacks systematic evaluations, standardized benchmarks, and well‑defined metrics for rigorously assessing the ASP‑coding capabilities of LLMs and, more generally, of any automatic approach.

\noindent\textbf{{Contributions.}} To fill this gap, we introduce \textbf{BLAST}, a benchmarking methodology specifically designed to assess the accuracy of LLMs in generating ASP code. 
%
BLAST provides a structured evaluation framework that measures program quality through two novel metrics specifically designed for ASP code generation. 
The methodology encompasses both evaluation \textit{breadth} (covering a diverse set of problems), and \textit{depth} (testing robustness across multiple natural-language paraphrases of each problem). Crucially, BLAST imposes no guidance or syntactic constraints on input/output formats for the evaluated models.
Moreover, existing procedures mostly rely on manual evaluations of the automated encoding outcomes~\citep{ijcai24-borroto,coppolilloLLASP}, resulting in time-consuming and potentially inconsistent assessments. 
In contrast, we propose a systematic approach that automates the evaluation process, reducing human intervention while improving scalability, reliability and reproducibility.

We assess BLAST via an experimental study that tackles ten well‑known graph problems from the ASP literature using a pool of eight state‑of‑the‑art LLMs.
%

The main contribution of the present work can be summarized as follows:
\begin{itemize}
\item 
    We release a benchmark dataset with ten graph-related problems to be encoded in ASP.
\item 
    We propose two novel evaluation methods to automatically assess the quality of the generated programs. Both approaches operate without imposing any constraints on the LLM’s input or output format, making the system robust to variations and renaming of input and output predicate symbols.
\item 
    We evaluate eight state-of-the-art LLMs, from five different families, by considering the original problem description along with paraphrased versions, to assess the model flexibility in interpreting human-like crafted prompts.
\end{itemize}

For the sake of reproducibility, the complete experimental setup, including code and evaluation data, is {made publicly available.\footnote{\url{https://anonymous.4open.science/r/LLMs-ASP-Benchmark-DFC3/}}}

\section{Related Work}\label{sec:related}

\paragraph{Automatic Code Synthesis.} The advantages of automating code synthesis are well-recognized in the literature~\citep{DBLP:journals/software/ErnstBM22,kalliamvakou2022research,DBLP:journals/corr/abs-2302-06590,DBLP:journals/jss/DakhelMNKDJ23}, and automated program composition tools now support nearly all mainstream programming languages~\citep{DBLP:journals/corr/abs-2107-03374}.
In this area, Large Language Models (LLMs) play a central role, and their performance has been extensively compared in the literature~\citep{evaluation-code,codet5}, while also the positive impact of fine-tuning models for code generation of imperative programming is established~\citep{llamoco}. 

In the context of declarative programming, a widely recognized research objective is to create tools that streamline and automate the development of ASP programs, thus bridging the gap between natural language specifications and ASP source code~\citep{DBLP:conf/bionlp/ErdemY09,DBLP:conf/inap/FangT17,DBLP:journals/tplp/Schwitter18,DBLP:journals/tplp/CarusoDMMR24}. 
Early proposals focused on automating the resolution of logic puzzles presented in simplified English by translating their descriptions into ASP~\citep{DBLP:conf/aaaifs/BaralD11}, employing $\lambda$-calculus and probabilistic combinatorial categorical grammars.
Later, several efforts have been spent in the development of Controlled Natural Languages (CNLs)~\citep{DBLP:journals/coling/Kuhn14} for ASP programs; 
CNLs represent subsets of full natural languages, featuring restricted grammar and vocabulary. 
Among them, the BIOQUERYCNL~\citep{DBLP:conf/bionlp/ErdemY09} defines the grammatical structure of a CNL and algorithm for transpiling queries into ASP; 
\citep{DBLP:conf/inap/FangT17} introduced a CNL approach for ASP that leverages LANA annotations, that was implemented in the SeaLion IDE. 
We note that \citep{DBLP:journals/tplp/Schwitter18} developed a CNL called PENG\textsuperscript{ASP} for specifying and verbalizing ASP programs, while
\citep{DBLP:journals/tplp/CarusoDMMR24} introduced CNL2ASP, an extensive publicly-available tool for converting controlled natural language into ASP programs. 

\paragraph{ASP and LLMs.} Powerful language tools such as LLMs have been exploited in conjunction with the ASP formalism in several ways. \citep{DBLP:conf/nips/NyeTTL21} introduced a dual-system model based on GPT-3, which generates semantic parsers from natural language sentences and integrates them with reasoning modules. In a similar vein, \citep{DBLP:conf/acl/YangI023} proposed that LLMs like GPT-3 can work as few-shot semantic parsers, transforming natural language into logical forms for ASP without necessitating distinct retraining for diverse question-answering tasks. 
\citep{DBLP:conf/kr/IshayY023} utilized LLMs with prompt engineering to obtain ASP solutions for logic puzzles, leveraging the logic puzzle dataset from~\citep{DBLP:conf/aaai/MitraB16}. 
\citep{DBLP:conf/lpnmr/KareemGBRR24} introduced LLM2LAS, which combines LLMs with ILASP~\citep{DBLP:journals/corr/abs-2005-00904} to learn commonsense knowledge from story-based Q\&A, enabling the system to generalize with minimal examples and effectively answer unseen questions.
Further, the effective use of LLMs in combination with ASP to perform a number of natural language understanding tasks has been explored and implemented in the STAR framework~\citep{reliable-nlu}.
{\cite{DBLP:journals/corr/abs-2502-09211} proposed NSGRAPH, an approach that integrates vision modules, LLM-based semantic parsing, and ASP reasoning for Visual Graph Question Answering, where LLMs extract ASP predicates from natural language questions that are then processed by an ASP solver.}
{Beyond ASP, \cite{DBLP:journals/jair/VoborilRS25} proposed StreamLLM, an approach that leverages LLMs to generate and validate streamlining constraints in MiniZinc models, achieving substantial speedups across multiple constraint satisfaction benchmarks.}
\citep{can-llms-solve-asp} introduced a benchmark focused on evaluating the ASP solving abilities of LLMs (i.e., entailment, verification and computation of answer sets). In contrast, BLAST focuses on the task of generating ASP code.
\citep{ijcai24-borroto} presented the NL2ASP tool, that constructs ASP programs from natural language specifications through a two-step architecture.
NL2ASP works by translating the NL specifications into CNL statements, which are then converted into ASP code using the CNL2ASP tool.
NL2ASP was implemented using well-known models for neural machine translation (NMT) and demonstrated promising performance. 
More recently,~\citep{coppolilloLLASP} provides a first evaluation of LLMs performance in encoding simple, core ASP programs, further showing that LLMs are not good at ASP coding, but fine-tuning can be beneficial. 
However, no benchmark or methodology for systematic assessment of LLM in the general ASP coding task is provided~\citep{coppolilloLLASP}.
{Building on the potential of fine-tuning, \citep{DBLP:journals/corr/abs-2512-17093} introduced an ASP-solver-in-the-loop framework that improves LLM-based ASP code generation for combinatorial problems by using direct solver feedback to curate data for supervised fine-tuning and to further improve robustness through a solver-guided search that includes best-of-N sampling.}


%
%
\paragraph{Evaluation Benchmarks.} A substantial body of literature 
provides benchmarking datasets and test suites tailored to evaluating LLMs on imperative (e.g., Python, C++, Java) and web-oriented programming languages (e.g., HTML, CSS, JavaScript) such as~\citep{xu2025webbenchllmcodebenchmark, 10.5555/3737916.3739082, 10.1145/3597503.3639219, DBLP:journals/corr/abs-2107-03374,ding2023crosscodeevaldiversemultilingualbenchmark, liu2023repobenchbenchmarkingrepositorylevelcode}. 
Comparatively, we argue that much less attention has been devoted to declarative programming paradigms. 

A first benchmark for automatic translation from natural language to ASP was introduced by \cite{ijcai24-borroto}. However, it primarily focuses on controlled natural language~\citep{DBLP:journals/tplp/CarusoDMMR24} and features input sentences with limited fluency.

{Recently, ~\cite{can-llms-solve-asp} proposed a methodology to assess whether state-of-the-art LLMs can {solve} programs coded in ASP, comparing their capabilities to standard solvers. However, their goal (computing answer sets with LLMs) is different from ours (testing LLMs coding ability). }

To the best of our knowledge, no existing benchmark enables systematic \textit{automatic} evaluation of LLMs ability to \textit{generate} ASP programs.
This work addresses this gap.  

\section{Preliminaries}\label{sec:preliminaries}

\paragraph{Large Language Model.}\label{subsec:llms} A Large Language Model (LLM) is a function $f$ that stochastically maps an input token sequence $\x = [x_1, \dots, x_n]$ to an output sequence $\y = [y_1, \dots, y_m]$, with $\x \in V^*$ and $\y \in W^*$, where $V$ and $W$ are token vocabularies. Formally, $f$ defines the conditional probability $\llmprob{\y}{\x}$, from which $\y$ is sampled, capturing the structure and semantics of natural language.
LLMs typically adopt the Transformer encoder-decoder architecture~\citep{attention-is-all}. The input tokens are first embedded into dense vectors encoding contextual and positional information. These embeddings pass through the Transformer: the encoder applies self-attention and feedforward layers to model dependencies within the input, while the decoder combines masked self-attention, encoder-decoder attention, and feedforward transformations to generate outputs token by token.
The decoder outputs a probability distribution over the output vocabulary via a linear layer followed by softmax:
\llmprob{y_i}{\x, \y_{:i-1}},
where $\y_{:i-1}$ is the prefix up to token $i{-}1$. Generation proceeds stepwise from this distribution. Architectural variants exist~\citep{comprehensive-overview,survey}, but their internal design is orthogonal to our focus.
%
LLMs are generally invoked via \textit{prompting}: a textual query (or \textit{prompt}) is provided as input to the model, which triggers content generation.

\paragraph{Answer Set Programming.}
Answer Set Programming (ASP) is a declarative formalism for Knowledge Representation and Reasoning~\citep{DBLP:journals/cacm/BrewkaET11,DBLP:journals/ngc/GelfondL91}.
In ASP, a problem is encoded as a finite set of logical rules, and solutions correspond to answer sets. Rather than presenting a formal introduction to the language, we provide a representative example to highlight its expressive power: the classical 3-colorability problem, which requires assigning one of three colors to each node of an undirected graph such that adjacent nodes receive different colors.

The following ASP encoding captures this specification. The input graph (nodes and edges) and the available colors are represented via facts. A choice rule assigns exactly one color to each node, and an integrity constraint eliminates colorings in which adjacent nodes share the same color:
\begin{footnotesize}
\begin{verbatim}
col(red). col(green). col(black). node(1..4).
edge(1,2). edge(4,2). edge(1,3).

1 { colored(X,C) : col(C) } 1 :- node(X).
:- edge(X,Y), colored(X,C), colored(Y,C).
\end{verbatim}
\end{footnotesize}

ASP also provides constructs for optimization via weak constraints, which allow specifying preferences among answer sets. For instance, to minimize the number of nodes assigned the color red: 

\begin{footnotesize}
\begin{flushleft}
    \verb|:|$\sim$ \verb|colored(X,red). [1@1,X]|
\end{flushleft}
\end{footnotesize}

This example illustrates the declarative nature and expressiveness of ASP in modeling complex combinatorial problems. For a rigorous introduction to the ASP language and its semantics, please refer to the aforesaid literature.
\section{BLAST Methodology}
\begin{figure}[]
    \centering
    \includegraphics[width=0.8\linewidth]{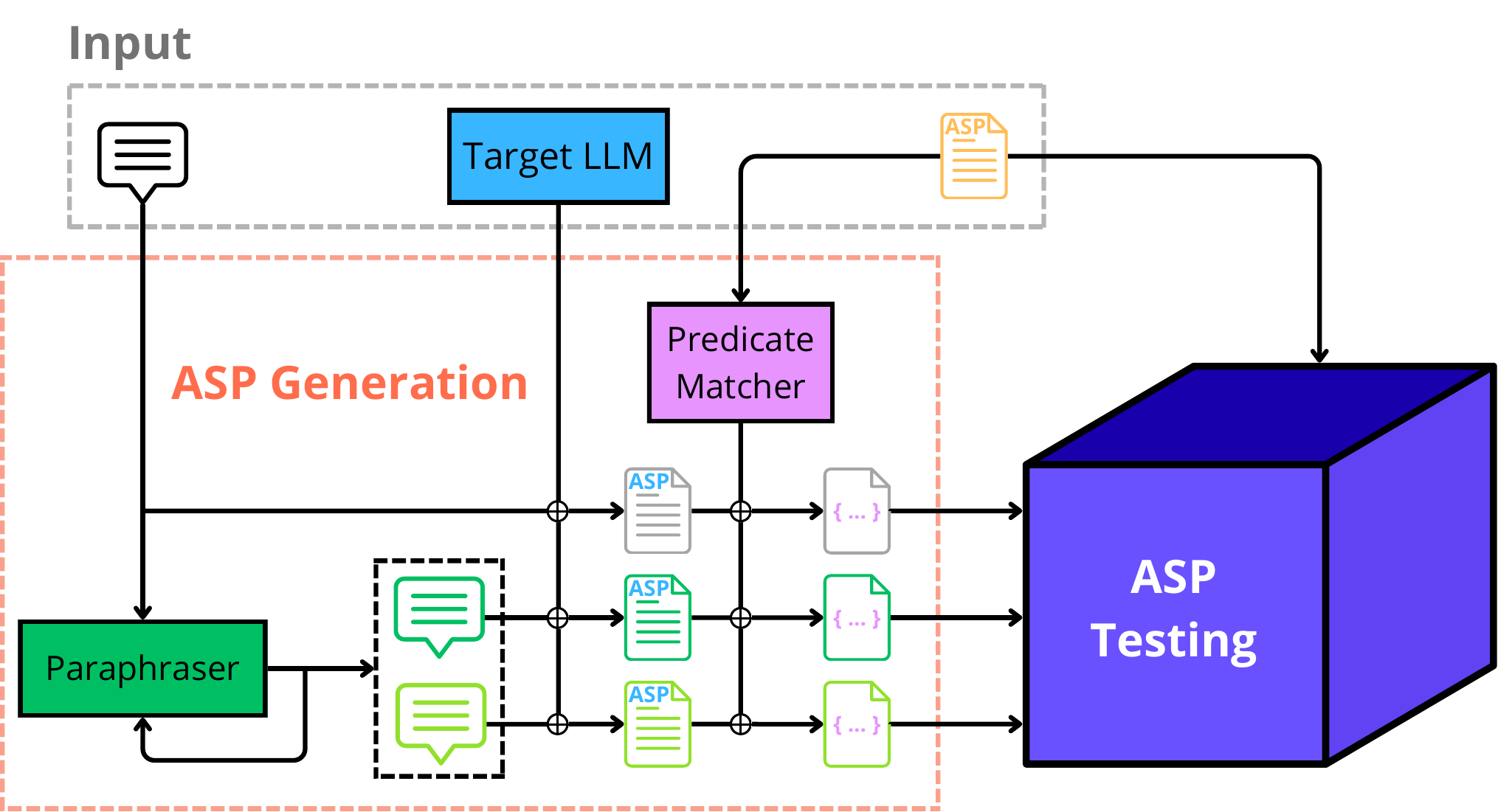}
    \caption{Scheme of the overall proposed framework. \textit{Input} consists of the \textit{textual specification} of the problem, the \textit{target LLM} to be evaluated, and the correct (\textit{gold}) \textit{ASP program}. The \textbf{ASP Generation} module comprises: the {\textit{paraphraser}}, an LLM which paraphrases the original problem description in more human-styled texts; and the \textit{predicate matcher}, an LLM which maps the predicates of the generated programs to the ones of the gold program. The predicate mappings and the gold encoding are finally provided to the \textbf{ASP Testing} module, which performs the evaluation.}
    \label{fig:approach}
\end{figure}

The core of the approach consists in testing the generation capabilities of a target LLM. For this, we devise a methodology consisting of two main components: 
\begin{enumerate*}[label=(\roman*),font=\itshape]
\item an \textbf{ASP Generation} framework that, starting from a specification in natural language, provides a set of encodings of the problem in ASP; 
\item an \textbf{ASP Testing} infrastructure aimed at evaluating the performance of the target LLM.
\end{enumerate*}
The overall scheme of the approach is detailed in Figure~\ref{fig:approach}. 

\subsection{ASP Generation}
The framework for generating ASP initiates with a natural language description of the problem intended for ASP encoding, alongside a \textit{gold program} that exemplifies the ASP-based solution. The generation process is driven by the following elements: \begin{enumerate*}[label=(\roman*),font=\itshape] \item the \textbf{target LLM}, serving as the generator under evaluation, in charge of encoding the problem into ASP; \item an LLM instructed to {\textbf{paraphrase}} the initial problem description to make it more human-readable; \item an LLM that semantically \textbf{matches} the predicates of the resulting program with those in the gold program. 
\end{enumerate*}
\paragraph{Target LLM.}
As an initial step in our framework, the LLM under evaluation is prompted with the problem description and asked to provide an encoding in ASP language. 
Box 1 provides an example of the provided prompt.
(Note that, this component can be replaced by any method generating ASP programs from natural language specifications.)

\mdfsetup{skipabove=5pt,skipbelow=5pt}
\begin{mdframed}[backgroundcolor=white!10,linecolor=gray!60!,roundcorner=0pt,linewidth=1pt,
rightline=false,
leftline=false] 
\begingroup
\fontsize{8.5pt}{10.5pt}\selectfont
\textbf{Box 1: Target LLM Prompt Example}
\\
\texttt{[PROBLEM DESCRIPTION]}
\\
\\
\textit{Write an ASP program which models this problem, along with facts that make it instantiable. Generate the program only, without any additional text.}
\endgroup
\label{box:tested_LLM_prompt}
\end{mdframed}


\paragraph{Predicate Matcher.}
It is important to remark that the target model is not given any instructions or constraints regarding the syntax, predicate names, or problem modeling to adopt. To normalize its output and enable comparison with the gold standard solution, we employ a further LLM, referred to as the Predicate Matcher. This model takes as input both the gold program and the encoding generated by the target model, and is prompted to perform a semantic match of the predicates used in the two programs. The result is returned as a Python dictionary mapping corresponding predicates. If no semantic similarity can be established between the programs, the model is instructed to return ``No semantic match''.
An example prompt of the predicate matcher is provided in Box 3, considering the Colorability problem.


\mdfsetup{skipabove=5pt,skipbelow=5pt}
\begin{mdframed}[backgroundcolor=white!10,linecolor=gray!60!,roundcorner=0pt,linewidth=1pt,
rightline=false,
leftline=false] 
\begingroup
\fontsize{8.5pt}{10.5pt}\selectfont
\textbf{Box 3: Predicate Matcher Prompt Example}
\\
\textit{Perform predicate matching on the following ASP programs based on the semantic similarity of the predicates:
\\
\\
ASP Instance 1:
\\}
\texttt{\scriptsize{
node(1). node(2). node(3). \\
edge(1,2).  edge(1,3).
edge(2,1). \\edge(2,3).
color(red). color(green). color(black).
\\
1 \{chosen(N,C) : color(C)\} 1 :- node(N).
\\
:- edge(U,V), U < V, chosen(U,C), chosen(V,C).}}
\\\\
\textit{ASP Instance 2:
\\}
\texttt{\scriptsize{
node(a).
node(b).
node(c).
\\
edge(a, b).
edge(b, c).
edge(c, d).
\\
colour(1).
colour(2).
colour(3).
\\
1 \{assign(N, C) : colour(C)\} 1 :- node(N).
\\
:- edge(X, Y), assign(X, C), assign(Y, C).}}\\
\\
\textit{If there is no semantic match between some predicates, output `No semantic match' only. Otherwise, produce the matches over all the predicates as a python dictionary. Do not generate any additional text.}
\endgroup
\label{box:matcher_prompt}
\end{mdframed}




\noindent
The generated matching dictionary is  the following:


\begin{flushleft}
\scriptsize{
\{`\texttt{node}':`\texttt{node}', `\texttt{edge}':`\texttt{edge}', `\texttt{color}':`\texttt{colour}', `\texttt{chosen}':`\texttt{assign}'\}}
\end{flushleft}

\noindent

\begin{figure}[!bht]
    \centering
    \includegraphics[width=0.65\linewidth]{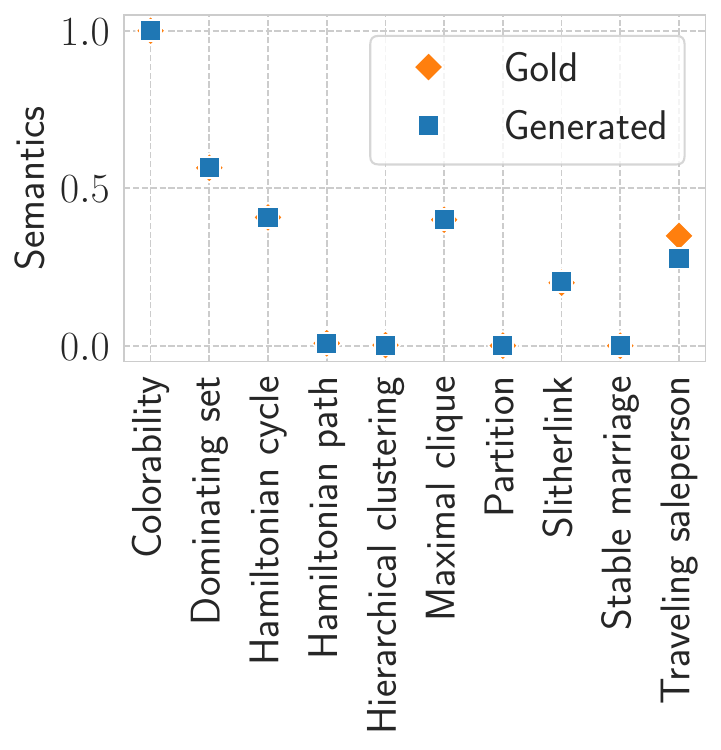}
    \caption{Semantic comparison on the considered problems computed by using the ``gold'' matching dictionary (manually crafted by authors) and the generated one. DeepSeek-R1 has been used as the target LLM.}
    \label{fig:gold-generated-match}
\end{figure}

We emphasize three crucial aspects of the predicate matcher component. First, the predicate matching phase is independent of encoding correctness: the matcher LLM only semantically associates predicates between programs, without assessing their validity. 
Second, the matcher does not semantically alter a correct program.
{In other words, the quality of the target LLM performance (that is, its ability to generate a correct ASP program from a natural language problem description) remains consistent regardless of whether a ``gold'' predicate matching dictionary (manually crafted by the authors) or a generated one is used (see Figure~\ref{fig:gold-generated-match}, where we use DeepSeek-R1 as the representative target LLM).}
The only exception is the Traveling Salesperson problem, where performance is slightly better with the gold matching.
By looking deeper into this problem, we realized that despite the better matching, the generated programs were still incorrect.
{Third, we remark that this result 
enables our methodology to achieve a high degree of generality, since it allows evaluating generated programs without imposing any prompting constraints on the LLMs.} 
Indeed, we deliberately avoid penalizing encodings that differ only at the syntactic level, through the use of alternative predicate names. For instance, adopting \texttt{reach(X)} instead of \texttt{reached(X)} to represent the same concept should not negatively affect the evaluation, provided that the program is semantically correct. By abstracting away from superficial naming differences, our approach focuses on assessing genuine modeling and coding ability rather than sensitivity to arbitrary lexical choices. 

\paragraph{{Paraphraser.}}
In addition to the technically detailed original problem statement, we further evaluate the adaptability of the target model to understand prompts that are formatted to be more relatable to humans. 
The goal is to evaluate the models not only in terms of \textit{breadth}, by addressing a diverse set of problems, but also in terms of \textit{depth}, by assessing their ability to adapt to less technical, more accessible textual descriptions intended for an ``ASP layman''. 
To achieve this, we use another LLM as a {\textit{paraphraser}} to generate two versions of the initial description.
An example prompt is reported in Box 2 in the {Supplementary Material}.

{The paraphrasing process is conducted in two phases. 
Initially, a first version is derived from the original problem statement. 
This version is then employed as a basis for producing a second one. 
Both versions are subsequently used to prompt the target LLM with alternative specifications.
Note that this phase is \textit{optional} in the proposed framework. While we study the impact of paraphrased textual descriptions on model performance as an additional analysis, the target LLM can still be evaluated using only the original encoding.}

\subsection{ASP Testing}\label{sect:testing}
To evaluate the ASP programs generated by the LLMs, we propose two methods: \textit{(i)} a \textbf{test-suite}-based approach and \textit{(ii)} a \textbf{model}-based approach.
Both methods are associated with proper metrics that allow us to measure a form of semantic distance from the gold program.
Both approaches assume that each problem has a fixed set of input and output predicates, corresponding to the ones of the gold program. 
This assumption is pragmatic, yet reasonably not too restrictive. 
Indeed, LLMs are expected to produce meaningful programs as written by a human; 
thus, syntactic variations of predicate names are very likely resolved by the predicate matcher. If not, we assume the  program is not correct.

\paragraph{Test-Suite-based Evaluation.} 
This evaluation method relies on the unit testing mechanism implemented in the ASP-WIDE tool~\citep{DBLP:journals/tplp/AmendolaMRB24}, which uses an annotation-based test specification language inspired by JUnit.
The objective is to create a test suite that thoroughly checks the program functionality.
A \textit{test suite} is a collection of test cases, each comprising assertions that verify a number of conditions on the expected program output.
Based on the test suite results, we compute the \textbf{test case accuracy} as the ratio of passed test cases over the total number. 

To ensure this evaluation method is effective, it is crucial to design a robust test suite over most of the program's expected behavior.
To this end, we developed a guideline for constructing meaningful test cases.
The first step involves creating test cases to check for unsatisfiability.
Specifically, given a set of input instances that violate the problem constraints, the program is expected to return no answer set for each instance. 
Next, we define test cases that verify the satisfiability of the problem constraints by checking if valid instances comply with them.
If needed, we further refine the test suite by adding test cases that validate the correctness of the program’s output atoms.
{For example, in the \textit{Colorability} problem, it should never occur that a node remains uncolored.
This is done by writing an ASP‑WIDE test case with the following assertion:}
\begin{small}
    \begin{verbatim}
@constraintForAll(constraint=":- node(U), 
    not chosenColor(U,_).")
    \end{verbatim}
\end{small}
{which ensures that every node $U$ is assigned at least one color.}
Finally, we add test cases to check whether the program produces the expected number of answer sets.
%

\color{black}
\paragraph{Model-based Testing.}
This method evaluates an ASP program $P_t$ based on its ability to reproduce the \textit{correct} (\textit{gold}) (stable) \textit{models} (GM) for a given set of instances provided via facts.
The core idea relies on the assumption that, given a gold program $P_g$ and a set of instances, the target program $P_t$ is considered correct if and only if it produces the \textit{ same} (stable) models as $P_g$ on those instances.
To formalize this, we define the set of \textit{target program models} (TPM) as the (stable) models of $P_t$, and the \textit{covered gold models} (CGM) as the intersection of TPM and GM, i.e., $\textrm{CGM} = \textrm{GM} \cap \textrm{TPM}$.

To quantify the accuracy of $P_t$, we use the standard (micro) F1-score~\citep{10.1145/3606367}, which combines precision and recall, defined as:
$$
    \textrm{Recall} = \frac{|\textrm{CGM}|}{|\textrm{GM}|} \quad \textrm{and} \quad \textrm{Precision} = \frac{|\textrm{CGM}|}{|\textrm{TPM}|},
$$
\noindent
The (micro) F1 score is then computed as the harmonic mean of precision and recall, providing a balanced measure of the program’s ability to reproduce the expected models.

\color{black}

Notably, we retrieve the CGM set with high efficiency by mean of the Algorithm~\ref{alg:gold_models}, which leverages an ASP solver.
Rather than relying on full answer set enumeration of the program under test, that might quickly become infeasible, we modify the ASP program to directly target specific solutions. 
In particular, we constrain the search space to gold models only, thereby avoiding the overhead of enumerating irrelevant answer sets. 
Clearly, we carefully crafted out instance set to be sure that the enumeration of gold program answers is instead feasible.

The algorithm takes as input the program to test $P_t$, a gold program $P_g$, and a set of instances $Ins$.
For each instance $I \in Ins$, we compute the set of gold models $AS_g$ by computing all answer sets of $P_g\cup I$ using an ASP solver (line~\ref{alg:golds}). 
Then, for each gold model $M_g \in AS_g$, we create a set of constraints aimed at forcing $M_g$ to be an answer set of $P_t$ (line~\ref{alg:constraints}).
Specifically, for each atom $a$ over the output predicates, if $a \in M_g$ then we add the constraint $\textit{:- not }a$; otherwise, we add the constraint $\textit{:- }a$.
\\

For the sake of clarity, let us consider the Colorability problem with $3$ nodes and $3$ colors and a gold model $M_g$
where node $1$ is colored as $purple$, node $2$ as $red$, and node $3$ as $brown$.
In this case, \texttt{create$\_$model$\_$constraints(M$\_$g)} returns the following constraints:
\begin{small}
    \begin{verbatim}
:- not chosenColor(3,brown).
:- not chosenColor(2,red).
:- not chosenColor(1,purple).
:- chosenColor(1,brown).
:- chosenColor(1,red).
:- chosenColor(2,brown).
:- chosenColor(2,purple).
:- chosenColor(3,red).
:- chosenColor(3,purple).
    \end{verbatim}
\end{small}
Intuitively, in the answer set $M_g$, node $1$ is colored as $purple$ (i.e., $chosenColor(1,purple)\in M_g$) and so $\textit{:- not }chosenColor(1,purple)$ is obtained. Conversely, node $1$ is colored neither $red$ nor $brown$, and so the constraints $\textit{:- }chosenColor(1,red)$ and $\textit{:- }chosenColor(1,brown)$ are obtained. Analogously, the remaining constraints are obtained for nodes $2$ and $3$.
Note that, if the gold program contains weak constraints then \texttt{create$\_$model$\_$constraints(M$\_$g)} does not return strong constraints as in the previous example but a set of weak constraints with cost 1 and a priority higher than the maximum priority used in $P_t$.
This ensures that the solver prioritizes the inclusion of these gold atoms in the resulting answer sets.
After obtaining the set of constraints $C_g$, the program $P_t \cup C_g\cup I$ is solved using the solver configured to return exactly one answer set $M$.
This operation is performed by the function \texttt{solve\_once} in line 9 of the algorithm.
Then, the gold model $M_g$ is considered \textit{covered} if $M_g$ exactly matches $M$. 
$M$ and $M_g$ exactly match if, for each atom $a$ over output predicates, it holds that $a\in M$ if and only if $a \in M_g$.

The next step is to identify the set of models generated by $P_t$ that differ from the gold models (i.e., wrong models).
{Note that this last step is necessary in order to properly calculate $TPM$.}
To this end, we first augment $P_t$ with the set of facts encoding the gold models in $AS_g$. 
To do this, we used the predicate \textit{trueInGold/2}, where the first term is an identifier (i.e., a positive integer) for a gold model $M_g\in AS_g$, and the second term is a true atom in $M_g$.  
Note that, such facts are obtained considering only atoms over the output predicates.
Then, the program $P_t$ is further augmented with additional rules that discard all those models of $P_t$ that exactly match one gold model. 
This augmented version of the program $P_t$ is returned by the function \texttt{aug\_program} and is denoted as $P_{ta}$.  
For example, let us consider again the Colorability problem and the gold model in which node $1$ is colored as $purple$, node $2$ as $red$, and node $3$ as $brown$. In this case the program $P_{ta}$ is obtained by adding to the program $P_t$ the rules reported in Listing~\ref{lst:listing}.

\begin{lstlisting}[caption=Rules and facts used to augment a program $P_t$ targeting the Colorability problem.,label=lst:listing,basicstyle=\scriptsize\ttfamily,numbers=left,breaklines=true,
    breakatwhitespace=true,
    columns=fullflexible,xleftmargin=1.7em]
trueInGold(1,chosenColor(2,red)).
trueInGold(1,chosenColor(1,purple)).
trueInGold(1,chosenColor(3,brown)).
...
mod(X) :- trueInGold(X,_).
trueInTested(chosenColor(X,Y)) :- chosenColor(X,Y).
smallerMG(M) :- trueInGold(M, X), not trueInTested(X).
smallerMt(M) :- mod(M), trueInTested(X), not trueInGold(M, X).
:- mod(M), not smallerMG(M), not smallerMt(M).
\end{lstlisting}

Rules in lines 1-3 in Listing~\ref{lst:listing} represent the facts encoding the considered gold model (here we reported only one for the sake of readability, but all gold models are considered).
The rule at line 6 is used to compute the true atoms in a model of $P_t$. 
Note that this rule is domain-specific, as the output predicate names vary across different domains.
Then, rules in lines 7-8 compute for each gold model, the differences with a candidate model of $P_t$.
Finally, the constraint at line 9 forces the program to discard a model if it matches a gold model.
%
Once the program $P_{ta}$ is obtained, the algorithm calls the solver to compute all models of $P_{ta}\cup I$ (i.e., \texttt{solve\_all}(P$_{ta}$$\cup$ I)).
Obtained models are considered wrong. 
Finally, the algorithm returns the set of gold models over all instances \textrm{GM}, the set of covered gold models $\textrm{CGM} \subseteq \textrm{GM}$, and the set of wrong models \textrm{WM}, which is disjoint from \textrm{GM}.
The value $|\textrm{TPM}|$ can then be calculated as the sum of the number of elements of \textrm{CGM} and \textrm{WM}.

\begin{algorithm}[t!]

\caption{Finding $P_t$ covered gold models and overall models.}
\label{alg:gold_models}
\textbf{Input}: A program $P_t$, a program $P_g$, a set of instances $Ins$.\\
\textbf{Output}: the sets of models $GM$, $CGM$, and $WM$. 
\begin{algorithmic}[1]
\STATE $GM = \emptyset$
\STATE $CGM = \emptyset$
\STATE $WM = \emptyset$
\FOR{$I \in Ins$} 
    \STATE $AS_g = \texttt{solve\_all}(P_g \cup I)$. 
    \label{alg:golds}
    \STATE $GM = GM \cup AS_g$. 
    \FOR {$M_g \in AS_g$}
        \STATE $C_g = \texttt{create\_model\_constraints}(M_g)$.
        \label{alg:constraints}
        \STATE $M = \texttt{solve\_once}(P_t \cup C_g \cup I)$.
        \IF {$M = M_g$}
            \STATE $CGM = CGM \cup \{M_g\}$
        \ENDIF
    \ENDFOR
    \STATE $P_{ta} = \texttt{aug\_program}(P_{t},AS_g)$.
    \STATE $AS_{w} = \texttt{solve\_all}(P_{ta} \cup I)$.
    \STATE $WM = WM \cup AS_{w}$
\ENDFOR
\STATE \textbf{return} $GM, CGM, WM$ 
\end{algorithmic}
\end{algorithm}

\paragraph{Discussion on Metrics.}
The introduced metrics are intended to assess how well an ASP program semantically models a given problem.
%
Among the two evaluation methods, the test-suite-based approach offers a more accurate measure of program correctness, as it involves white-box testing. 
This type of testing enables detailed inspection of a program internal behavior, rather than relying solely on its output models.
However, the test-suite-based approach has some limitations; 
in fact, it requires a human expert to design the test cases, and the reliability of the final assessment depends heavily on the quality of the test suite. 
In contrast, the model-based approach offers a higher level of automation, as human intervention is limited to providing input instances and the gold program.
This method, however, makes it more difficult to know the reasons why a program may be incorrect, which limits potential efforts to improve it.
Indeed, we might inspect the results and discover some answer set is missing or wrong, but there is no clue on why it happens. On the  other hand, specific test cases are devised with the aim at testing a specific ability, and thus are more connected to a reason for the failure.



\section{Experimental Evaluation}
\textcolor{black}{We present our experimental evaluation, which tackles the following research questions:
\begin{itemize}[leftmargin=1cm]
    \item[\textbf{RQ1:}] Are state-of-the-art LLMs \textit{effective} in ASP coding, under both syntactic and semantic perspectives? 
    \item[\textbf{RQ2:}] How does {\textit{paraphrasing}} impact on performance?
    \item[\textbf{RQ3:}] How do the proposed semantic metrics \textit{compare}, in terms of accuracy and efficiency?  
\end{itemize}
}
\subsection{Setup and Validation Protocol}
\paragraph{Dataset} To evaluate the proposed approach, we created a dataset comprising the following ten graph-related problems: \textbf{Colorability}, \textbf{(Connected) Dominating Set}, \textbf{Hamiltonian Cycle}, \textbf{Hamiltonian Path}, \textbf{Hierarchical Clustering}, \textbf{Maximal Clique}, \textbf{(Graph) Partitioning}, \textbf{Slitherlink}, \textbf{(Strong) Stable Marriage}, and \textbf{Traveling Salesman}. 
We selected this set of tasks since they are standard problems used in many ASP competitions~\citep{DBLP:journals/jair/GebserMR17,DBLP:journals/tplp/CalimeriIR14,DBLP:conf/lpnmr/AlvianoCCDDIKKOPPRRSSSWX13}. 
Each problem includes a formal natural language description and the corresponding gold program.
For each problem, we provide the related test suite and the input instances required by the evaluation methods outlined earlier.
\textcolor{black}{In the Supplementary Material}, we provide the problem descriptions and details regarding the test suite design.
\paragraph{Test Suite Validation.} 
To ensure the quality of our test suite, we draw inspiration from Oetsch et al. \citeyearpar{DBLP:conf/kr/OetschPPST12}, who empirically demonstrated that the small-scope hypothesis of traditional testing is also applicable to ASP programs, i.e., many errors can be detected by testing a program against inputs involving only a small number of objects.
Following Amendola et al. \citep{DBLP:journals/tplp/AmendolaMRB24}, we employ a mutation-based testing strategy specifically designed for ASP~\citep{DBLP:conf/kr/OetschPPST12}. Given an ASP program, the tool randomly applies a set of modifications to its rules, generating \textit{mutants}, altered versions of the original program that introduce subtle errors.
For each gold program, we generate 15 mutants and execute the corresponding test suite on them using our selected input instances. After each execution, we iteratively refine the test suite and the input instances until none of the mutants achieves a perfect test score (i.e., $1.0$).
This iterative process ensures that the test suite is effective in detecting deviations from the expected program behavior.
%
\paragraph{Tested Models.}
For the experimental evaluation, we adopt eight state-of-the-art LLMs from five distinct families: \textbf{o4-mini}
~\citep{o4mini}, \textbf{Claude-3.5-Sonnet}
~\citep{Claude3S}, \textbf{DeepSeek-R1}
~\citep{deepseek2025r1}, \textbf{Gemini-1.5-Pro}
~\citep{geminiteam2024gemini15unlockingmultimodal}, \textbf{Gemini-2.5-Flash}
~\citep{comanici2025gemini25pushingfrontier}, \textbf{GPT-4o}
~\citep{openai2024gpt4ocard}, \textbf{GPT-5}~\citep{OpenAI2025GPT5}, and \textbf{LLaMa3.3-70B}
~\citep{grattafiori2024llama3herdmodels}. 
    
\paragraph{Settings.} All tested models have been invoked via the dedicated APIs of OpenAI, Anthropic, Google, and DeepSeek, with a temperature of $0.7$ (when allowed). 
In the case of LLaMa3.3-70B, we used a self-hosted instance using the Ollama tool.
The LLM-related and ASP-testing code has been implemented in Python. 
The interaction with the ASP-WIDE tool (developed in Spring Boot) was done through a RESTful API endpoint.
The experiments have been conducted on a server equipped with an Intel(R) Xeon(R) CPU E7-8880 v4 @ 2.20GHz, 512GB of RAM, and running Debian 13 (6.1.0-32-amd64).
Additional information and experimental results are provided as Supplementary Material. {The experimental code is made publicly available.}\footnote{\url{https://anonymous.4open.science/r/LLMs-ASP-Benchmark-DFC3/}}

\begin{figure}[t]
    \centering
    \includegraphics[width=0.85\linewidth]{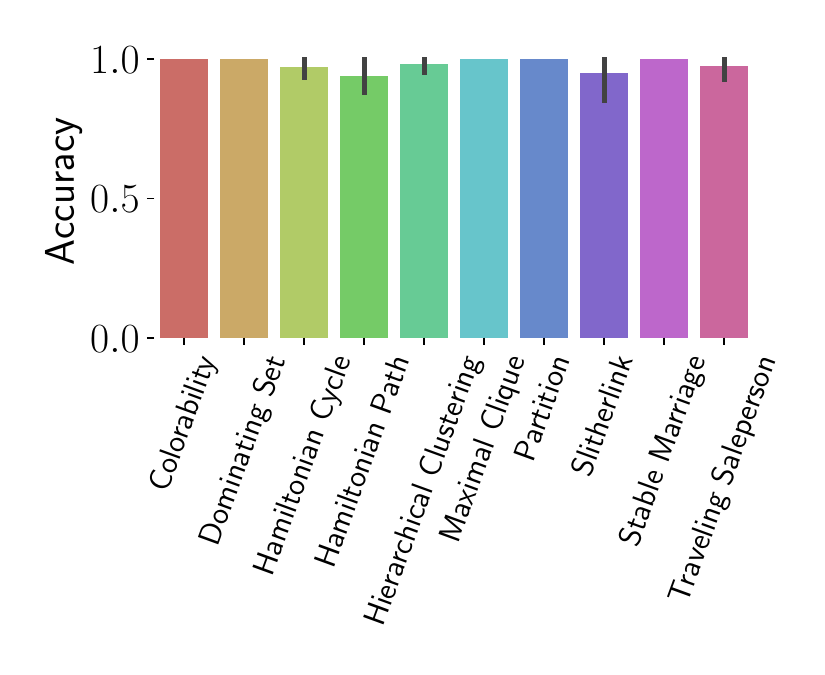}
    \caption{Performance of GPT-4o as predicate matcher. Error bars represent confidence intervals on ten different runs.}
    \label{fig:entity-matcher-accuracy}
\end{figure}
\subsection{Results}

\paragraph{Predicate Matcher Accuracy.}  We first assess the predicate matcher accuracy in mapping predicates from the two input programs semantically. 
To achieve this, we manually verify the matches, assigning each generated dictionary a score ranging from $[0,1]$. 
Specifically, we count the correct associations and average them over the total number of predicates. 
For each {problem}, we create ten separate matches to ensure statistical significance and report the average as the ultimate performance score. 
Figure~\ref{fig:entity-matcher-accuracy} presents the results using GPT-4o as the predicate matcher.
The accuracy of the model on this task comes as no surprise, given its well-documented abilities to perform text summarization and rephrasing~\citep{adams-etal-2023-sparse, zhou2025gptmodelsfollowhuman}; thus, we opt for GPT-4o also as a {paraphraser}.

\begin{figure}[t]
    \centering
    \includegraphics[width=\linewidth]{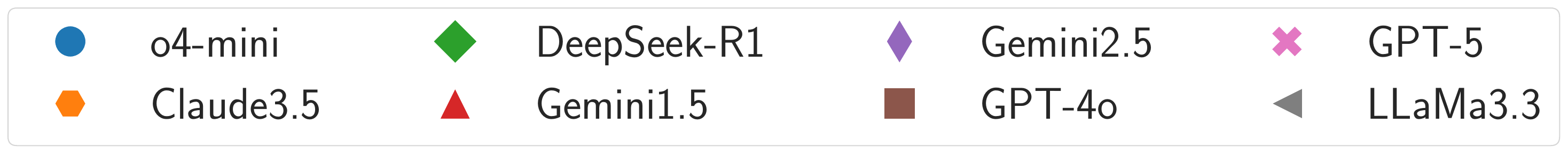}

    \includegraphics[width=0.9\linewidth]{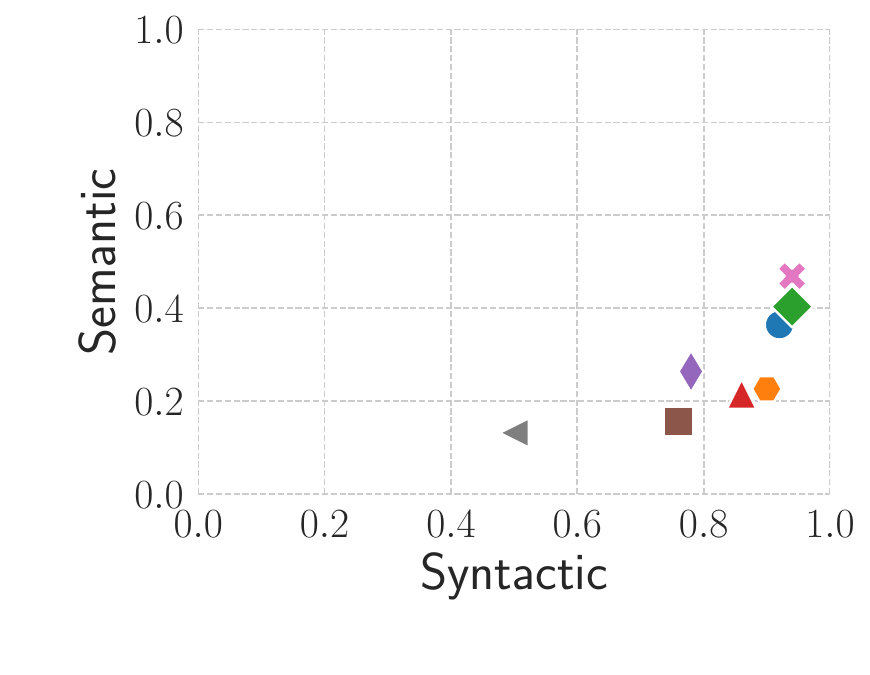}
    \caption{Trade-off of the evaluated LLMs between syntactic (X-axis) and semantic accuracy (Y-axis). The semantic score is computed according to the test suite-based method\protect\footnotemark[7].
    Results are averaged across all considered problems.}
    \label{fig:syntactic-semantic-scatterplot}
\end{figure}

\begin{figure}[t]
    \centering
    \includegraphics[width=0.6\linewidth]{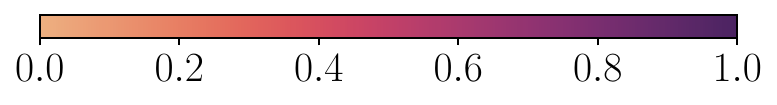}
    \\
    \begin{subfigure}[b]{0.5555\linewidth}
        
    \includegraphics[width=\linewidth]{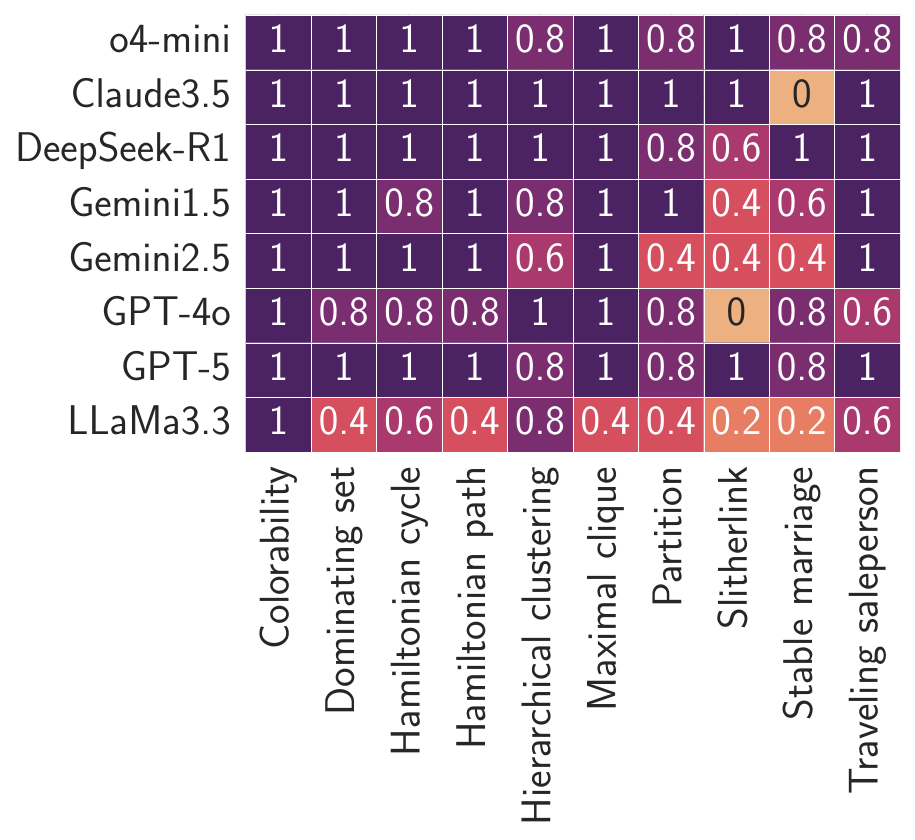}
    \caption{\textbf{Syntactic}}
    \end{subfigure}
    \begin{subfigure}[b]{0.42\linewidth}
        
    \includegraphics[width=\linewidth]{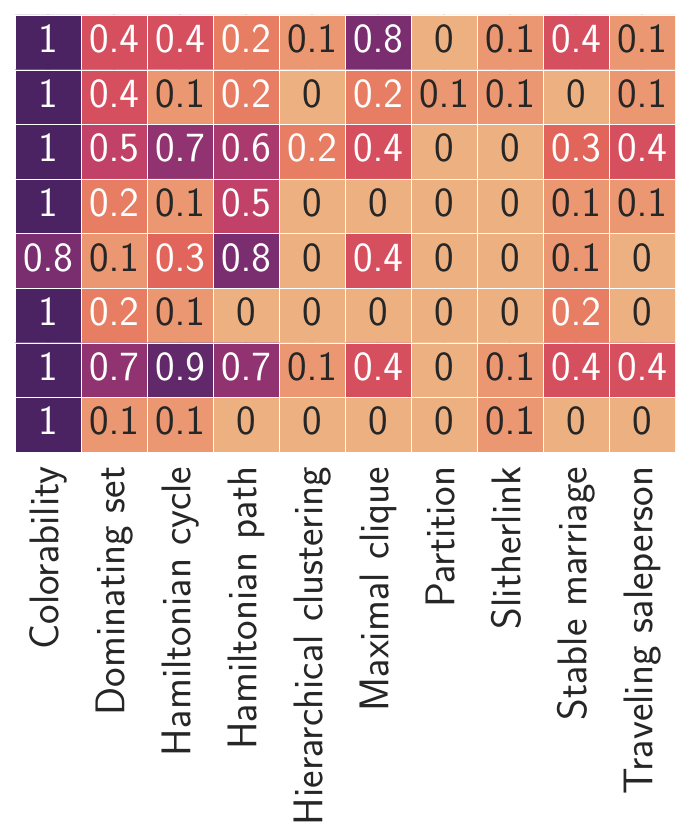}
    \caption{\textbf{Semantic}}
    \end{subfigure}
    \caption{Syntactic and semantic performance obtained by the evaluated LLMs across the considered problems. The test suite-based metric is used as the semantic reference\protect\footnotemark[7]. Each value represents the average of five different runs.}
    \label{fig:syntactic-semantic-heatmaps}
\end{figure}

\paragraph{LLMs Performance \textbf{(RQ1)}.} Further, we assess the effectiveness of the evaluated models concerning both syntactic syntax and semantic accuracy. Figure~\ref{fig:syntactic-semantic-scatterplot} shows the average results for all examined problems, whereas Figure~\ref{fig:syntactic-semantic-heatmaps} displays scores for each individual problem. At this stage, we concentrate solely on the initial problem statements, with more in-depth analyses discussed subsequently. Here, semantic accuracy is determined using test suite-based evaluation\footnotetext[7]{For Maximal Clique, we use only the model-based metric due to ASPWIDE run-time issues.}. In comparison, syntactic correctness is evaluated by checking whether the input program results in any syntax errors during processing by the solver. To ensure results are statistically robust, each experiment is conducted by calling the target LLM five times, with outcomes then averaged.

Two considerations can be made. First, as depicted in Figure~\ref{fig:syntactic-semantic-scatterplot}, while most LLMs show an impressive accuracy under the syntactic perspective (with the notable exception of LLaMa3.3), they perform poorly in terms of semantics. Secondly, Figure~\ref{fig:syntactic-semantic-heatmaps} shows that, while simple problems such as Colorability or Dominating Set are easier to address, almost all LLMs fall short in encoding more complex tasks (e.g., Partition or Slitherlink). 
%
This suggests that state-of-the-art LLMs still struggle to effectively encode declarative paradigms, such as Answer Set Programming. 

\begin{figure}[b]
    \centering
    \includegraphics[width=1\linewidth]{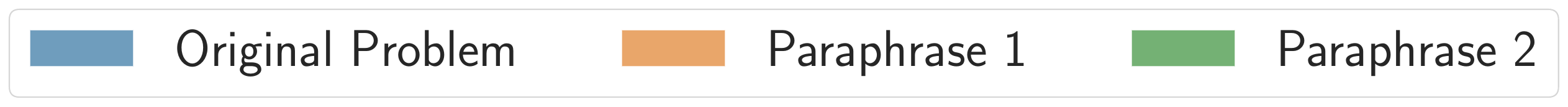}
    \includegraphics[width=1\linewidth]{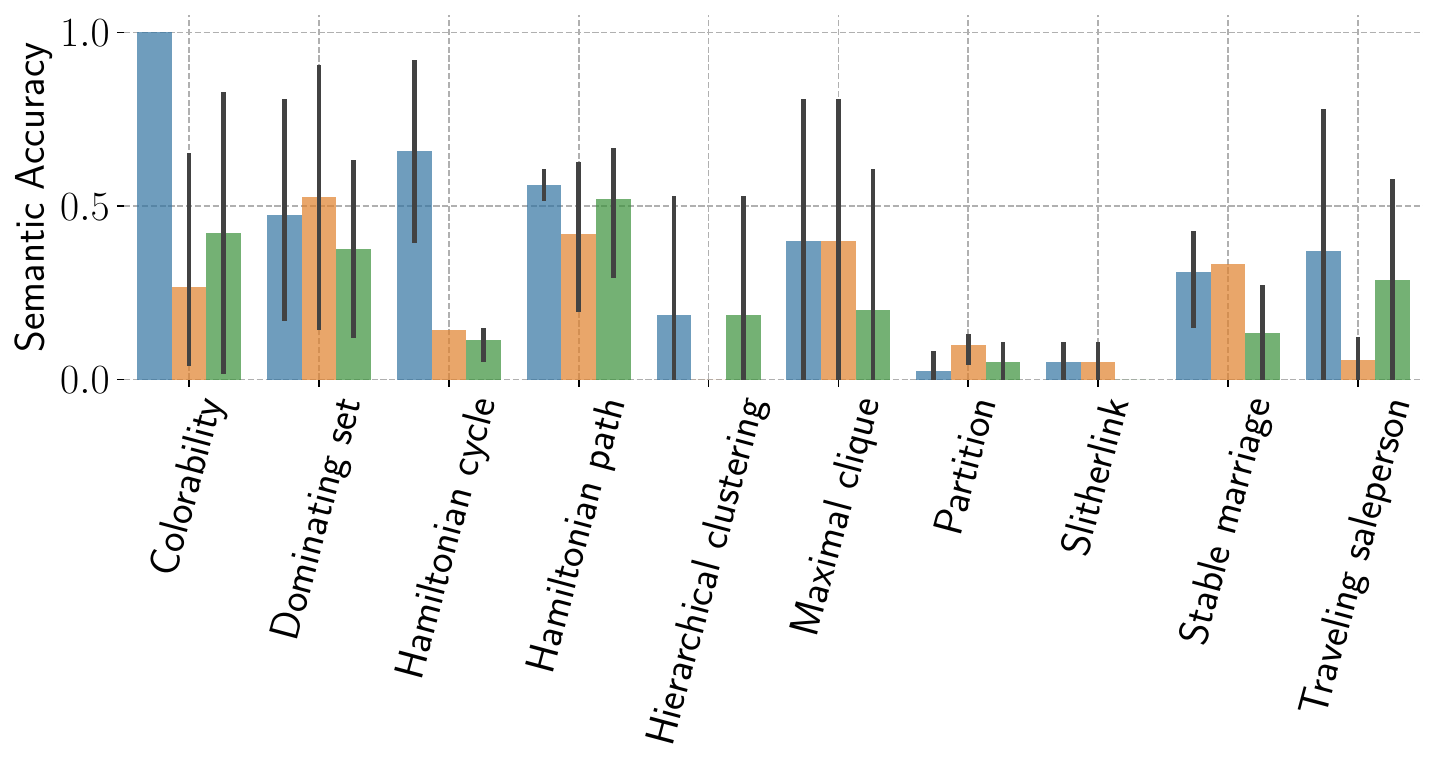}
    \caption{Comparison of the semantic accuracy computed via the test-suite when problem {paraphrasings} are applied\protect\footnotemark[7]. DeepSeek-R1 is used as the reference LLM. Bars depict the average across five different runs, while error bars represent confidence intervals.}
    \label{fig:semantic-simplifications}
\end{figure}

\paragraph{{Paraphrasing} Impact \textbf{(RQ2)}.} Further, we want to assess whether a more natural, human-like style of the textual description might help in increasing performance. To this aim, we select DeepSeek-R1 as the representative LLM candidate, since it achieves one of the best syntactic-semantic trade-off while being the cheapest to probe. We compare its semantic accuracy when problem {paraphrasings} are applied.
Figure~\ref{fig:semantic-simplifications}
depicts the test-suite-based evaluation score\footnotemark[7], averaged over five runs, along with the confidence intervals. We see that, in most cases, performance is leveraged when the original problem description is used, despite some exceptions where the first {paraphrasing} slightly improves the semantic accuracy of the model (notably, with Dominating set, Partition, Stable marriage).

\paragraph{{LLM Error Analysis.}}
{For completeness, we summarize the most common error patterns observed in syntactically valid but semantically incorrect programs. A frequent issue is the omission of constraints. For example, in the Hamiltonian Path problem, LLMs often fail to enforce that exactly one edge enters and one edge leaves each node, which is required to guarantee a proper path. In Stable Marriage, models sometimes omit bijectivity constraints ensuring that each man is matched with exactly one woman and vice versa. In optimization tasks involving weak constraints, mistakes in the specification of weights or priorities are also common. Another recurring problem concerns incorrect predicate signatures, such as wrong arity or mismatched arguments in input atoms. In some cases, additional constraints are introduced without a clear correspondence to the problem description; while not always harmful, they reveal a tendency to produce plausible but unjustified rules. Overall, LLMs tend to generate encodings that look reasonable at a superficial level, yet the formal rigor of ASP exposes subtle inaccuracies. Although such drafts may assist human programmers, our results indicate that fully automatic generation of correct ASP encodings remains challenging for current state-of-the-art models.}

\begin{figure}[t]
    \centering
    \includegraphics[width=\linewidth]{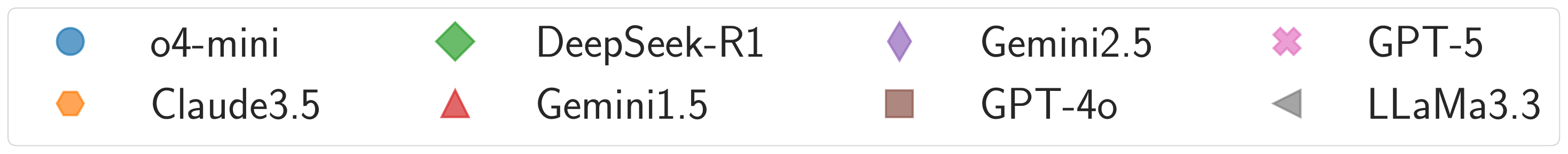}

    \includegraphics[width=\linewidth]{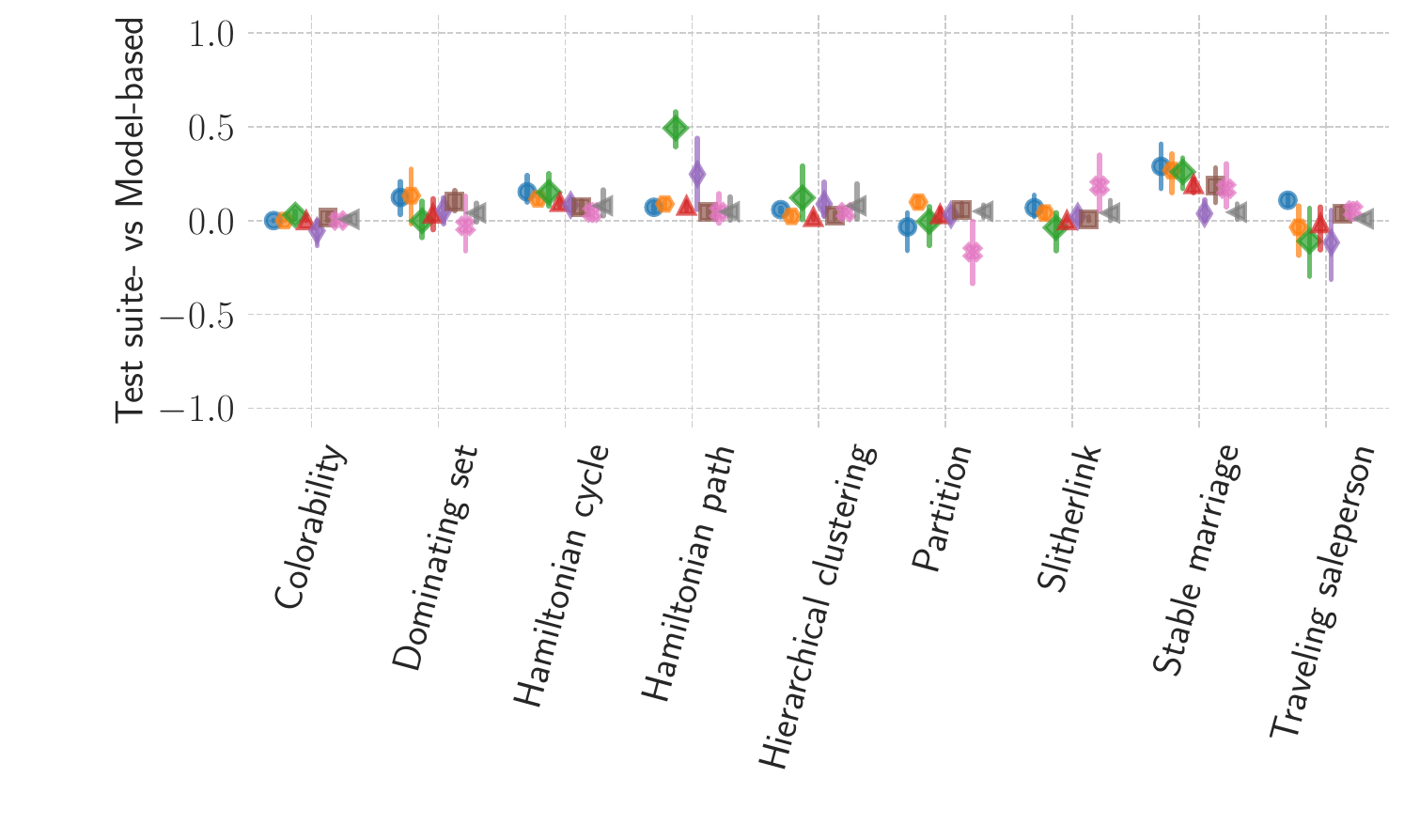}
    \caption{Comparison between the two proposed semantic metrics\protect\footnotemark[8]. DeepSeek-R1 is the reference LLM. Values represent the difference between test-suite- and model-based scores, averaged across five runs, while lines indicate confidence intervals.}
    \label{fig:comparing-semantic-metrics}
\end{figure}

\begin{figure}[t]
    \centering
    \includegraphics[width=0.75\linewidth]{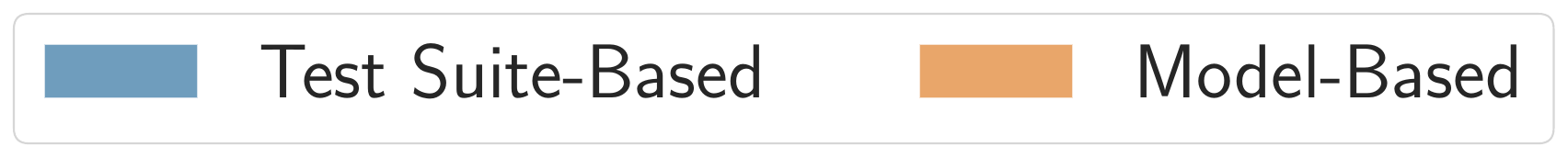}

    \includegraphics[width=0.9\linewidth]{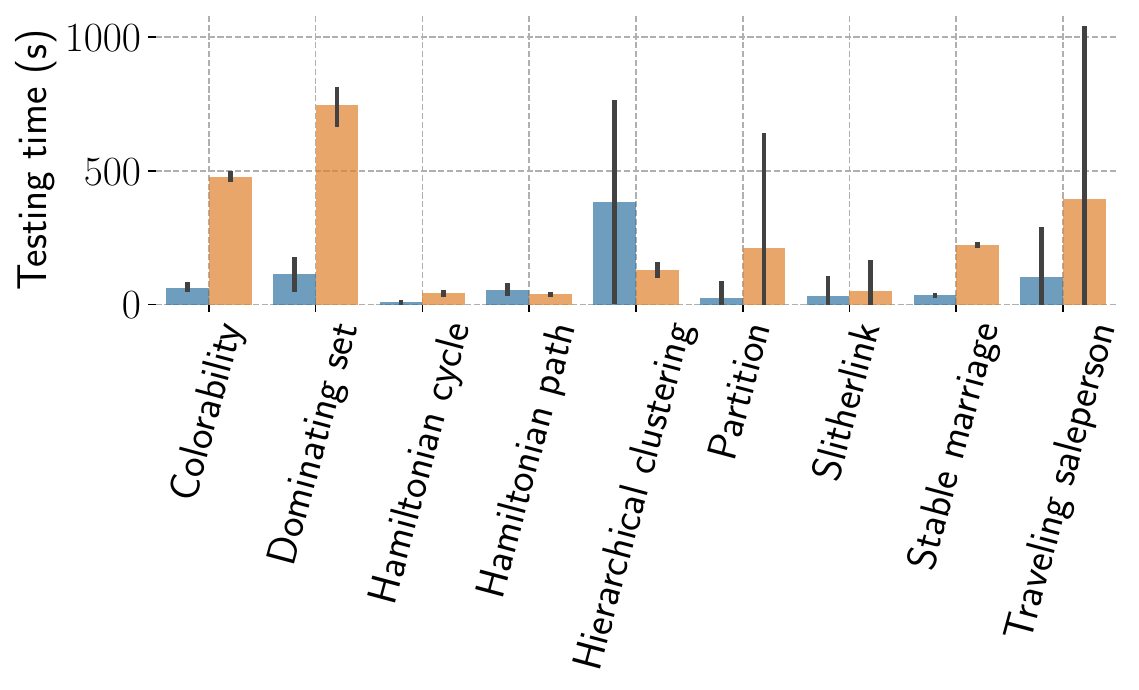}
    \caption{Comparison between the two semantic metrics in terms of testing times (expressed in seconds). DeepSeek-R1 is used as the reference LLM. Bars represent the average across five runs, error bars indicate confidence intervals.}
    \label{fig:testing-times}
\end{figure}

\paragraph{Test-suite- vs Model-based \textbf{(RQ3)}.} Next, we compare the two proposed semantic metrics, both in terms of accuracy (Figure~\ref{fig:comparing-semantic-metrics}) and efficiency (Figure~\ref{fig:testing-times}). Specifically, Figure~\ref{fig:comparing-semantic-metrics} shows the difference in performance computed via the test-suite- and the model-based metric
Therefore, a positive (resp., negative) score means that the performance is higher if the test-suite-based metric (resp., the model-based metric) is considered. Again, results are obtained by averaging five distinct runs, while the bars represent confidence intervals. 
We see that, across most problems and evaluated LLMs, the two metrics align. This alignment is also confirmed by the Pearson correlation coefficient $\rho=0.865$. 

We finally compare the two metrics in terms of efficiency. Figure~\ref{fig:testing-times}
depicts the testing times (in seconds) across the considered problems, using DeepSeek-R1 as the reference LLM. In all problems (except for Hierarchical clustering), the test-suite-based metric significantly reduces the computation time, resulting in the most efficient metric.
(More details in the Supplementary Material.)




\section{Conclusions and Future Work}
Although LLMs are known to excel at language understanding and code synthesis, their ability to generate correct ASP programs has remained comparatively underexplored. Research on the automatic composition of ASP code is growing; however, evaluations often rely on manual inspection or impose constraints on the LLM’s input/output format, making performance sensitive to superficial variations such as predicate renaming. As a consequence, the field still lacks systematic standardized benchmarks and rigorous metrics for assessing the ASP coding capabilities of LLMs and, more broadly, of automatic program coding approaches.

In this paper, we addressed this gap by introducing \textbf{BLAST}, the first dedicated benchmarking methodology and accompanying dataset specifically designed to rigorously evaluate the accuracy of LLMs in generating ASP programs. Along with BLAST, we implement and release an ad-hoc benchmark, two testing methodologies and related correctness metrics.
The proposed metrics measure the semantic distance between the generated encoding and the reference specification, without imposing constraints on predicate names or LLMs prompt structure. 
We show that the two measures consistently align in their assessment of correctness, while remaining complementary: one is more transparent and interpretable, at the cost of requiring human intervention, whereas the other is fully automatic but less directly explainable. Building on BLAST, we conduct an extensive empirical study of eight state-of-the-art LLMs from five different model families, evaluating their ability to encode ten well-known graph problems in ASP starting from textual descriptions.
Experiments reveal that: (\textit{i}) although LLMs often achieve high syntactic accuracy, they struggle with semantic correctness; (\textit{ii}) their performance degrades significantly as task complexity increases; and (\textit{iii}) rephrasing the problem specification does not lead to consistent improvements.

\medskip
As future work, we plan to address current limitations and explore several directions for extension. Although our evaluation adopts zero-shot prompting to ensure a uniform and controlled comparison, more advanced strategies, such as few-shot prompting, prompt tuning, or retrieval-augmented generation (RAG), deserve systematic investigation, and could lead to measure better performance of LLMs in generating ASP code. 
Moreover, our benchmark focuses on graph problems that are standard ASP system testing. 
Extending the benchmark to additional domains would further strengthen its scope and representativeness. 
Finally, although BLAST here evaluates general-purpose LLMs, our methodology is model-agnostic and could also be employed to evaluate specialized techniques for ASP encoding.


\section*{{GenAI Statement}}
{This paper has been fully conceptualized, designed, and written by the human authors. LLMs were used to proofread the paper and identify some typos and style issues, the authors claim the full ownership of the text.
Importantly, LLM are used as integral components of the proposed framework, as detailed in the manuscript, and are subject solely to experimental evaluation. }

\section*{{Acknowledgments}}
This work has been partially funded by MUR on D.M.\ 351/2022, PNRR Ricerca, CUP H23C22000440007; and partially supported by Moneying-Plus project (CUP J29I24001790005) selected within the framework of the PR FESR – FSE Calabria 2021/2027 and implemented with the support of the Italian State and the Calabria Region.
The work has also been partially supported by PN RIC project ASVIN ``Assistente Virtuale Intelligente di Negozio'' (CUP B29J24000200005) and the PNRR project FAIR - Future AI Research (PE00000013), Spoke 9 -- Green-aware AI, under the NRRP MUR program funded by the ``NextGenerationEU''.

\clearpage

\bibliographystyle{kr}
\bibliography{ref}

@inproceedings{attention-is-all,
 author = {Vaswani, Ashish and Shazeer, Noam and Parmar, Niki and Uszkoreit, Jakob and Jones, Llion and Gomez, Aidan N and Kaiser, \L ukasz and Polosukhin, Illia},
 booktitle = {Advances in Neural Information Processing Systems},
 editor = {I. Guyon and U. Von Luxburg and S. Bengio and H. Wallach and R. Fergus and S. Vishwanathan and R. Garnett},
 pages = {},
 publisher = {Curran Associates, Inc.},
 title = {Attention is All you Need},
 volume = {30},
 year = {2017}
}

@misc{DBLP:journals/corr/abs-2404-07751,
  author       = {Pavel Smirnov and
                  Frank Joublin and
                  Antonello Ceravola and
                  Michael Gienger},
  title        = {Generating consistent {PDDL} domains with Large Language Models},
  journal      = {CoRR},
  volume       = {abs/2404.07751},
  year         = {2024}
}

@misc{DBLP:journals/tmlr/ValmeekamSGK25,
  author       = {Karthik Valmeekam and
                  Kaya Stechly and
                  Atharva Gundawar and
                  Subbarao Kambhampati},
  title        = {A Systematic Evaluation of the Planning and Scheduling Abilities of
                  the Reasoning Model o1},
  journal      = {Trans. Mach. Learn. Res.},
  volume       = {2025},
  year         = {2025}
}

@inproceedings{DBLP:conf/esws/LippolisSKZCGBN25,
  author       = {Anna Sofia Lippolis and
                  Mohammad Javad Saeedizade and
                  Robin Keskis{\"{a}}rkk{\"{a}} and
                  Sara Zuppiroli and
                  Miguel Ceriani and
                  Aldo Gangemi and
                  Eva Blomqvist and
                  Andrea Giovanni Nuzzolese},
  title        = {Ontology Generation Using Large Language Models},
  booktitle    = {{ESWC} {(1)}},
  series       = {Lecture Notes in Computer Science},
  volume       = {15718},
  pages        = {321--341},
  publisher    = {Springer},
  year         = {2025}
}

@misc{DBLP:journals/ki/FalknerFSTT18,
  author       = {Andreas A. Falkner and
                  Gerhard Friedrich and
                  Konstantin Schekotihin and
                  Richard Taupe and
                  Erich Christian Teppan},
  title        = {Industrial Applications of Answer Set Programming},
  journal      = {K{\"{u}}nstliche Intell.},
  volume       = {32},
  number       = {2-3},
  pages        = {165--176},
  year         = {2018}
}

@misc{comprehensive-overview,
      title={A Comprehensive Overview of Large Language Models}, 
      author={Humza Naveed and Asad Ullah Khan and Shi Qiu and Muhammad Saqib and Saeed Anwar and Muhammad Usman and Naveed Akhtar and Nick Barnes and Ajmal Mian},
      year={2024},
      eprint={2307.06435},
      archivePrefix={arXiv},
      primaryClass={cs.CL}
}

@misc{survey,
  author={Raiaan, Mohaimenul Azam Khan and Mukta, Md. Saddam Hossain and Fatema, Kaniz and Fahad, Nur Mohammad and Sakib, Sadman and Mim, Most Marufatul Jannat and Ahmad, Jubaer and Ali, Mohammed Eunus and Azam, Sami},
  journal={IEEE Access}, 
  title={A Review on Large Language Models: Architectures, Applications, Taxonomies, Open Issues and Challenges}, 
  year={2024},
  volume={12},
  number={},
  pages={26839-26874},
  doi={10.1109/ACCESS.2024.3365742}}

@misc{DBLP:journals/cacm/BrewkaET11,
  author       = {Gerhard Brewka and
                  Thomas Eiter and
                  Miroslaw Truszczynski},
  title        = {Answer set programming at a glance},
  journal      = {Commun. {ACM}},
  volume       = {54},
  number       = {12},
  pages        = {92--103},
  year         = {2011}
}

@misc{DBLP:journals/ngc/GelfondL91,
  author       = {Michael Gelfond and
                  Vladimir Lifschitz},
  title        = {Classical Negation in Logic Programs and Disjunctive Databases},
  journal      = {New Gener. Comput.},
  volume       = {9},
  number       = {3/4},
  pages        = {365--386},
  year         = {1991}
}

@misc{o4mini,
  author       = {OpenAI},
  title        = {Introducing OpenAI o3 and o4-mini},
  year         = {2025},
  url          = {https://openai.com/index/introducing-o3-and-o4-mini/}
}

@misc{deepseek2025r1,
  author       = {DeepSeek},
  title        = {DeepSeek-R1: Incentivizing Reasoning Capability in LLMs via Reinforcement Learning},
  year         = {2025},
  url          = {https://arxiv.org/abs/2501.12948}
}

@inproceedings{coppolilloLLASP,
author = {Coppolillo, Erica and Calimeri, Francesco and Manco, Giuseppe and Perri, Simona and Ricca, Francesco},
title = {LLASP: fine-tuning large language models for answer set programming},
year = {2024},
doi = {10.24963/kr.2024/78},
booktitle = {Proceedings of the 21st International Conference on Principles of Knowledge Representation and Reasoning},
series = {KR '24}
}

@misc{xu2025webbenchllmcodebenchmark,
      title={Web-Bench: A LLM Code Benchmark Based on Web Standards and Frameworks}, 
      author={Kai Xu and YiWei Mao and XinYi Guan and ZiLong Feng},
      year={2025},
      eprint={2505.07473},
      archivePrefix={arXiv},
      primaryClass={cs.AI},
      url={https://arxiv.org/abs/2505.07473}, 
}

@inproceedings{10.5555/3737916.3739082,
author = {Shah, Nidhish and Genc, Zulkuf and Araci, Dogu},
title = {StackEval: benchmarking LLMs in coding assistance},
year = {2025},
booktitle = {Proceedings of the 38th International Conference on Neural Information Processing Systems},
numpages = {19},
series = {NIPS '24}
}

@inproceedings{10.1145/3597503.3639219,
author = {Du, Xueying and Liu, Mingwei and Wang, Kaixin and Wang, Hanlin and Liu, Junwei and Chen, Yixuan and Feng, Jiayi and Sha, Chaofeng and Peng, Xin and Lou, Yiling},
title = {Evaluating Large Language Models in Class-Level Code Generation},
year = {2024},
doi = {10.1145/3597503.3639219},
booktitle = {Proceedings of the IEEE/ACM 46th International Conference on Software Engineering},
series = {ICSE '24}
}

@misc{ding2023crosscodeevaldiversemultilingualbenchmark,
      title={CrossCodeEval: A Diverse and Multilingual Benchmark for Cross-File Code Completion}, 
      author={Yangruibo Ding and Zijian Wang and Wasi Uddin Ahmad and Hantian Ding and Ming Tan and Nihal Jain and Murali Krishna Ramanathan and Ramesh Nallapati and Parminder Bhatia and Dan Roth and Bing Xiang},
      year={2023},
      eprint={2310.11248},
      archivePrefix={arXiv},
      primaryClass={cs.LG},
      url={https://arxiv.org/abs/2310.11248}, 
}

@misc{liu2023repobenchbenchmarkingrepositorylevelcode,
      title={RepoBench: Benchmarking Repository-Level Code Auto-Completion Systems}, 
      author={Tianyang Liu and Canwen Xu and Julian McAuley},
      year={2023},
      eprint={2306.03091},
      archivePrefix={arXiv},
      primaryClass={cs.CL},
      url={https://arxiv.org/abs/2306.03091}, 
}

@misc{DBLP:journals/software/ErnstBM22,
  author       = {Neil A. Ernst and
                  Gabriele Bavota},
  title        = {AI-Driven Development Is Here: Should You Worry?},
  journal      = {{IEEE} Softw.},
  volume       = {39},
  number       = {2},
  pages        = {106--110},
  year         = {2022}
}

@misc{kalliamvakou2022research,
  title={Research: quantifying GitHub Copilot’s impact on developer productivity and happiness},
  author={Kalliamvakou, Eirini},
  journal={The GitHub Blog},
  year={2022}
}

@misc{DBLP:journals/corr/abs-2302-06590,
  author       = {Sida Peng and
                  Eirini Kalliamvakou and
                  Peter Cihon and
                  Mert Demirer},
  title        = {The Impact of {AI} on Developer Productivity: Evidence from GitHub
                  Copilot},
  journal      = {CoRR},
  volume       = {abs/2302.06590},
  year         = {2023}
}

@misc{DBLP:journals/jss/DakhelMNKDJ23,
  author       = {Arghavan Moradi Dakhel and
                  Vahid Majdinasab and
                  Amin Nikanjam and
                  Foutse Khomh and
                  Michel C. Desmarais and
                  Zhen Ming (Jack) Jiang},
  title        = {GitHub Copilot {AI} pair programmer: Asset or Liability?},
  journal      = {J. Syst. Softw.},
  volume       = {203},
  pages        = {111734},
  year         = {2023}
}

@inproceedings{leveraging-llms,
author = {Ishay, Adam and Yang, Zhun and Lee, Joohyung},
title = {Leveraging large language models to generate answer set programs},
year = {2023},
isbn = {978-1-956792-02-7},
url = {https://doi.org/10.24963/kr.2023/37},
doi = {10.24963/kr.2023/37},
booktitle = {Proceedings of the 20th International Conference on Principles of Knowledge Representation and Reasoning},
articleno = {37},
numpages = {10},
location = {Rhodes, Greece},
series = {KR '23}
}

@misc{reliable-nlu,
   title={Reliable Natural Language Understanding with Large Language Models and Answer Set Programming},
   volume={385},
   ISSN={2075-2180},
   url={http://dx.doi.org/10.4204/EPTCS.385.27},
   DOI={10.4204/eptcs.385.27},
   journal={Electronic Proceedings in Theoretical Computer Science},
   publisher={Open Publishing Association},
   author={Rajasekharan, Abhiramon and Zeng, Yankai and Padalkar, Parth and Gupta, Gopal},
   year={2023},
   month=sep, pages={274–287} 
}

@inproceedings{evaluation-code,
author = {Xu, Frank F. and Alon, Uri and Neubig, Graham and Hellendoorn, Vincent Josua},
title = {A systematic evaluation of large language models of code},
year = {2022},
isbn = {9781450392730},
publisher = {Association for Computing Machinery},
address = {New York, NY, USA},
url = {https://doi.org/10.1145/3520312.3534862},
doi = {10.1145/3520312.3534862},
booktitle = {Proceedings of the 6th ACM SIGPLAN International Symposium on Machine Programming},
pages = {1–10},
numpages = {10},
location = {San Diego, CA, USA},
series = {MAPS 2022}
}

@misc{codet5,
      title={CodeT5+: Open Code Large Language Models for Code Understanding and Generation}, 
      author={Yue Wang and Hung Le and Akhilesh Deepak Gotmare and Nghi D. Q. Bui and Junnan Li and Steven C. H. Hoi},
      year={2023},
      eprint={2305.07922},
      archivePrefix={arXiv},
      primaryClass={cs.CL}
}

@misc{llamoco,
      title={LLaMoCo: Instruction Tuning of Large Language Models for Optimization Code Generation}, 
      author={Zeyuan Ma and Hongshu Guo and Jiacheng Chen and Guojun Peng and Zhiguang Cao and Yining Ma and Yue-Jiao Gong},
      year={2024},
      eprint={2403.01131},
      archivePrefix={arXiv},
      primaryClass={math.OC}
}

@misc{DBLP:journals/corr/abs-2107-03374,
  author       = {Mark Chen and
                  Jerry Tworek and
                  Heewoo Jun et al.
                  },
  title        = {Evaluating Large Language Models Trained on Code},
  journal      = {CoRR},
  volume       = {abs/2107.03374},
  year         = {2021}
}

@inproceedings{DBLP:conf/bionlp/ErdemY09,
  author       = {Esra Erdem and
                  Reyyan Yeniterzi},
  title        = {Transforming Controlled Natural Language Biomedical Queries into Answer
                  Set Programs},
  booktitle    = {BioNLP@HLT-NAACL},
  pages        = {117--124},
  publisher    = {ACL},
  year         = {2009}
}

@misc{DBLP:journals/tplp/Schwitter18,
  author       = {Rolf Schwitter},
  title        = {Specifying and Verbalising Answer Set Programs in Controlled Natural
                  Language},
  journal      = {Theory Pract. Log. Program.},
  volume       = {18},
  number       = {3-4},
  pages        = {691--705},
  year         = {2018}
}

@inproceedings{DBLP:conf/inap/FangT17,
  author       = {Min Fang and
                  Hans Tompits},
  title        = {An Approach for Representing Answer Sets in Natural Language},
  booktitle    = {{DECLARE}},
  series       = {LNCS},
  volume       = {10997},
  pages        = {115--131},
  publisher    = {Springer},
  year         = {2017}
}

@misc{DBLP:journals/tplp/CarusoDMMR24,
  author       = {Simone Caruso and
                  Carmine Dodaro and
                  Marco Maratea and
                  Marco Mochi and
                  Francesco Riccio},
  title        = {{CNL2ASP:} Converting Controlled Natural Language Sentences into {ASP}},
  journal      = {Theory Pract. Log. Program.},
  volume       = {24},
  number       = {2},
  pages        = {196--226},
  year         = {2024}
}

@inproceedings{ijcai24-borroto,
  author       = {Manuel Borroto and
                  Irfan Kareem and
                  Francesco Ricca},
  title        = {Towards Automatic Composition of {ASP} Programs from Natural Language
                  Specifications},
  booktitle    = {{IJCAI}},
  pages        = {to appear},
  publisher    = {ijcai.org},
  year         = {2024},
  journal      = {CoRR},
  volume       = {abs/2403.04541}
}

@misc{DBLP:journals/aim/ErdemGL16,
  author       = {Esra Erdem and
                  Michael Gelfond and
                  Nicola Leone},
  title        = {Applications of Answer Set Programming},
  journal      = {{AI} Mag.},
  volume       = {37},
  number       = {3},
  pages        = {53--68},
  year         = {2016}
}

@inproceedings{DBLP:conf/lpnmr/AlvianoCDFLPRVZ17,
  author       = {Mario Alviano and
                  Francesco Calimeri and
                  Carmine Dodaro and
                  Davide Fusc{\`{a}} and
                  Nicola Leone and
                  Simona Perri and
                  Francesco Ricca and
                  Pierfrancesco Veltri and
                  Jessica Zangari},
  title        = {The {ASP} System {DLV2}},
  booktitle    = {{LPNMR}},
  series       = {LNCS},
  volume       = {10377},
  pages        = {215--221},
  publisher    = {Springer},
  year         = {2017}
}

@inproceedings{DBLP:conf/iclp/GebserKKOSW16,
  author       = {Martin Gebser and
                  Roland Kaminski and
                  Benjamin Kaufmann and
                  Max Ostrowski and
                  Torsten Schaub and
                  Philipp Wanko},
  title        = {Theory Solving Made Easy with Clingo 5},
  booktitle    = {{ICLP} (Technical Communications)},
  series       = {OASIcs},
  volume       = {52},
  pages        = {2:1--2:15},
  publisher    = {Schloss Dagstuhl - Leibniz-Zentrum f{\"{u}}r Informatik},
  year         = {2016}
}

@misc{DBLP:journals/coling/Kuhn14,
  author       = {Tobias Kuhn},
  title        = {A Survey and Classification of Controlled Natural Languages},
  journal      = {Comput. Linguistics},
  volume       = {40},
  number       = {1},
  pages        = {121--170},
  year         = {2014}
}

@inproceedings{DBLP:conf/aaaifs/BaralD11,
  author       = {Chitta Baral and
                  Juraj Dzifcak},
  title        = {Solving Puzzles Described in English by Automated Translation to Answer
                  Set Programming and Learning How To Do That Translation},
  booktitle    = {{AAAI} Fall Symposium: Advances in Cognitive Systems},
  series       = {{AAAI} Technical Report},
  volume       = {{FS-11-01}},
  publisher    = {{AAAI}},
  year         = {2011}
}

@inproceedings{DBLP:conf/aaai/MitraB16,
  author       = {Arindam Mitra and
                  Chitta Baral},
  title        = {Addressing a Question Answering Challenge by Combining Statistical
                  Methods with Inductive Rule Learning and Reasoning},
  booktitle    = {{AAAI}},
  pages        = {2779--2785},
  publisher    = {{AAAI} Press},
  year         = {2016}
}

@inproceedings{DBLP:conf/nips/NyeTTL21,
  author       = {Maxwell I. Nye and
                  Michael Henry Tessler and
                  Joshua B. Tenenbaum and
                  Brenden M. Lake},
  title        = {Improving Coherence and Consistency in Neural Sequence Models with
                  Dual-System, Neuro-Symbolic Reasoning},
  booktitle    = {NeurIPS},
  pages        = {25192--25204},
  year         = {2021}
}

@inproceedings{DBLP:conf/acl/YangI023,
  author       = {Zhun Yang and
                  Adam Ishay and
                  Joohyung Lee},
  title        = {Coupling Large Language Models with Logic Programming for Robust and
                  General Reasoning from Text},
  booktitle    = {{ACL} (Findings)},
  pages        = {5186--5219},
  publisher    = {ACL},
  year         = {2023}
}

@misc{DBLP:journals/tplp/BusoniuOPST13,
  author       = {Paula{-}Andra Busoniu and
                  Johannes Oetsch and
                  J{\"{o}}rg P{\"{u}}hrer and
                  Peter Skocovsky and
                  Hans Tompits},
  title        = {SeaLion: An eclipse-based {IDE} for answer-set programming with advanced
                  debugging support},
  journal      = {Theory Pract. Log. Program.},
  volume       = {13},
  number       = {4-5},
  pages        = {657--673},
  year         = {2013}
}

@inproceedings{DBLP:conf/lpnmr/FebbraroRR11,
  author       = {Onofrio Febbraro and
                  Kristian Reale and
                  Francesco Ricca},
  title        = {{ASPIDE:} Integrated Development Environment for Answer Set Programming},
  booktitle    = {{LPNMR}},
  series       = {LNCS},
  volume       = {6645},
  pages        = {317--330},
  publisher    = {Springer},
  year         = {2011}
}

@inproceedings{DBLP:conf/iclp/AlvianoCR23,
  author       = {Mario Alviano and
                  Davide Cirimele and
                  Luis Angel Rodriguez Reiners},
  title        = {Introducing {ASP} recipes and {ASP} Chef},
  booktitle    = {{ICLP} Workshops},
  series       = {{CEUR} WP},
  volume       = {3437},
  publisher    = {CEUR-WS.org},
  year         = {2023}
}

@misc{DBLP:journals/corr/abs-2303-10118,
  author       = {Susana Hahn and
                  Orkunt Sabuncu and
                  Torsten Schaub and
                  Tobias Stolzmann},
  title        = {Clingraph: {A} System for ASP-based Visualization},
  journal      = {CoRR},
  volume       = {abs/2303.10118},
  year         = {2023}
}

@misc{DBLP:journals/tplp/AmendolaMRB24,
  author       = {Giovanni Amendola and
                  Giuseppe Mazzotta and
                  Francesco Ricca and
                  Tobias Berei},
  title        = {Unit Testing in {ASP} Revisited: Language and Test-Driven Development
                  Environment},
  journal      = {{Theory Pract. Log. Program.}},
  volume       = {24},
  number       = {6},
  pages        = {1078--1108},
  year         = {2024}
}

@inproceedings{DBLP:conf/kr/OetschPPST12,
  author       = {Johannes Oetsch and
                  Michael Prischink and
                  J{\"{o}}rg P{\"{u}}hrer and
                  Martin Schwengerer and
                  Hans Tompits},
  title        = {On the Small-Scope Hypothesis for Testing Answer-Set Programs},
  booktitle    = {{KR}},
  publisher    = {{AAAI} Press},
  year         = {2012}
}

@misc{zhao2025surveylargelanguagemodels,
      title={A Survey of Large Language Models}, 
      author={Wayne Xin Zhao and Kun Zhou and Junyi Li et al.},
      year={2025},
      eprint={2303.18223},
      archivePrefix={arXiv},
      primaryClass={cs.CL},
      url={https://arxiv.org/abs/2303.18223}, 
}

@misc{qin2024largelanguagemodelsmeet,
      title={Large Language Models Meet NLP: A Survey}, 
      author={Libo Qin and Qiguang Chen and Xiachong Feng and Yang Wu and Yongheng Zhang and Yinghui Li and Min Li and Wanxiang Che and Philip S. Yu},
      year={2024},
      eprint={2405.12819},
      archivePrefix={arXiv},
      primaryClass={cs.CL},
      url={https://arxiv.org/abs/2405.12819}, 
}

@inproceedings{ou-etal-2024-dialogbench,
    title = "{D}ialog{B}ench: Evaluating {LLM}s as Human-like Dialogue Systems",
    author = "Ou, Jiao  and
      Lu, Junda  and
      Liu, Che  and
      Tang, Yihong  and
      Zhang, Fuzheng  and
      Zhang, Di  and
      Gai, Kun",
    booktitle = "Proceedings of the 2024 Conference of the North American Chapter of the Association for Computational Linguistics: Human Language Technologies (Volume 1: Long Papers)",
    month = jun,
    year = "2024",
    address = "Mexico City, Mexico",
    publisher = "Association for Computational Linguistics",
    url = "https://aclanthology.org/2024.naacl-long.341/",
    doi = "10.18653/v1/2024.naacl-long.341",
}

@misc{jiang2024surveylargelanguagemodels,
      title={A Survey on Large Language Models for Code Generation}, 
      author={Juyong Jiang and Fan Wang and Jiasi Shen and Sungju Kim and Sunghun Kim},
      year={2024},
      eprint={2406.00515},
      archivePrefix={arXiv},
      primaryClass={cs.CL},
      url={https://arxiv.org/abs/2406.00515}, 
}

@misc{geminiteam2024gemini15unlockingmultimodal,
      title={Gemini 1.5: Unlocking multimodal understanding across millions of tokens of context}, 
      author={Gemini Team and Petko Georgiev et al.},
      year={2024},
      eprint={2403.05530},
      archivePrefix={arXiv},
      primaryClass={cs.CL},
      url={https://arxiv.org/abs/2403.05530}, 
}

@misc{openai2024gpt4ocard,
      title={GPT-4o System Card}, 
      author={OpenAI},
      year={2024},
      eprint={2410.21276},
      archivePrefix={arXiv},
      primaryClass={cs.CL},
      url={https://arxiv.org/abs/2410.21276}, 
}

@misc{grattafiori2024llama3herdmodels,
      title={The Llama 3 Herd of Models}, 
      author={Aaron Grattafiori and Abhimanyu Dubey et al.},
      year={2024},
      eprint={2407.21783},
      archivePrefix={arXiv},
      primaryClass={cs.AI},
      url={https://arxiv.org/abs/2407.21783}, 
}

@inproceedings{DBLP:conf/lpnmr/KareemGBRR24,
  author       = {Irfan Kareem and
                  Katie Gallagher and
                  Manuel A. Borroto and
                  Francesco Ricca and
                  Alessandra Russo},
  title        = {Using Learning from Answer Sets for Robust Question Answering with
                  {LLM}},
  booktitle    = {{LPNMR}},
  series       = {Lecture Notes in Computer Science},
  volume       = {15245},
  pages        = {112--125},
  publisher    = {Springer},
  year         = {2024}
}

@misc{DBLP:journals/corr/abs-2005-00904,
  author       = {Mark Law and
                  Alessandra Russo and
                  Krysia Broda},
  title        = {The {ILASP} system for Inductive Learning of Answer Set Programs},
  journal      = {CoRR},
  volume       = {abs/2005.00904},
  year         = {2020}
}

@inproceedings{Claude3S,
  title={Claude 3.5 Sonnet Model Card Addendum},
  author={Anthropic},
year={2024},
  url={https://api.semanticscholar.org/CorpusID:270667923}
}

@inproceedings{adams-etal-2023-sparse,
    title = "From Sparse to Dense: {GPT}-4 Summarization with Chain of Density Prompting",
    author = "Adams, Griffin  and
      Fabbri, Alex  and
      Ladhak, Faisal  and
      Lehman, Eric  and
      Elhadad, No{\'e}mie",
    booktitle = "Proceedings of the 4th New Frontiers in Summarization Workshop",
    year = "2023",
    publisher = "Association for Computational Linguistics",
    doi = "10.18653/v1/2023.newsum-1.7",
    pages = "68--74",
}

@misc{zhou2025gptmodelsfollowhuman,
      title={Can GPT models Follow Human Summarization Guidelines? A Study for Targeted Communication Goals}, 
      author={Yongxin Zhou and Fabien Ringeval and François Portet},
      year={2025},
      eprint={2310.16810},
      archivePrefix={arXiv},
      primaryClass={cs.CL},
      url={https://arxiv.org/abs/2310.16810}, 
}

@misc{10.1145/3606367,
author = {Christen, Peter and Hand, David J. and Kirielle, Nishadi},
title = {A Review of the F-Measure: Its History, Properties, Criticism, and Alternatives},
year = {2023},
issue_date = {March 2024},
publisher = {Association for Computing Machinery},
address = {New York, NY, USA},
volume = {56},
number = {3},
issn = {0360-0300},
doi = {10.1145/3606367},
journal = {ACM Comput. Surv.},
}

@inproceedings{DBLP:conf/kr/IshayY023,
  author       = {Adam Ishay and
                  Zhun Yang and
                  Joohyung Lee},
  title        = {Leveraging Large Language Models to Generate Answer Set Programs},
  booktitle    = {{KR}},
  pages        = {374--383},
  year         = {2023}
}

@misc{DBLP:journals/jair/GebserMR17,
  author       = {Martin Gebser and
                  Marco Maratea and
                  Francesco Ricca},
  title        = {The Sixth Answer Set Programming Competition},
  journal      = {J. Artif. Intell. Res.},
  volume       = {60},
  pages        = {41--95},
  year         = {2017}
}

@inproceedings{DBLP:conf/lpnmr/AlvianoCCDDIKKOPPRRSSSWX13,
  author       = {Mario Alviano and
                  Francesco Calimeri and
                  G{\"{u}}nther Charwat and
                  Minh Dao{-}Tran and
                  Carmine Dodaro and
                  Giovambattista Ianni and
                  Thomas Krennwallner and
                  Martin Kronegger and
                  Johannes Oetsch and
                  Andreas Pfandler and
                  J{\"{o}}rg P{\"{u}}hrer and
                  Christoph Redl and
                  Francesco Ricca and
                  Patrik Schneider and
                  Martin Schwengerer and
                  Lara Katharina Spendier and
                  Johannes Peter Wallner and
                  Guohui Xiao},
  title        = {The Fourth Answer Set Programming Competition: Preliminary Report},
  booktitle    = {{LPNMR}},
  series       = {Lecture Notes in Computer Science},
  volume       = {8148},
  pages        = {42--53},
  publisher    = {Springer},
  year         = {2013}
}

@misc{DBLP:journals/tplp/CalimeriIR14,
  author       = {Francesco Calimeri and
                  Giovambattista Ianni and
                  Francesco Ricca},
  title        = {The third open answer set programming competition},
  journal      = {Theory Pract. Log. Program.},
  volume       = {14},
  number       = {1},
  pages        = {117--135},
  year         = {2014}
}

@misc{DBLP:journals/tplp/GebserMR20,
  author       = {Martin Gebser and
                  Marco Maratea and
                  Francesco Ricca},
  title        = {The Seventh Answer Set Programming Competition: Design and Results},
  journal      = {Theory Pract. Log. Program.},
  volume       = {20},
  number       = {2},
  pages        = {176--204},
  year         = {2020}
}

@misc{OpenAI2025GPT5,
  author = {{OpenAI}},
  title = {Introducing GPT-5},
  year = {2025},
  url = {https://openai.com/index/introducing-gpt-5/},
}

@inproceedings{can-llms-solve-asp,
author = {Ren, Lin and Xiao, Guohui and Qi, Guilin and Geng, Yishuai and Xue, Haohan},
title = {Can LLMs solve ASP problems? insights from a benchmarking study},
year = {2025},
isbn = {978-1-956792-08-9},
url = {https://doi.org/10.24963/kr.2025/60},
doi = {10.24963/kr.2025/60},
booktitle = {Proceedings of the 22nd International Conference on Principles of Knowledge Representation and Reasoning},
articleno = {60},
numpages = {11},
location = {Melbourne, Australia},
series = {KR '25}
}

@misc{comanici2025gemini25pushingfrontier,
      title={Gemini 2.5: Pushing the Frontier with Advanced Reasoning, Multimodality, Long Context, and Next Generation Agentic Capabilities}, 
      author={Gheorghe Comanici and Eric Bieber and Mike Schaekermann et al.},
      year={2025},
      eprint={2507.06261},
      archivePrefix={arXiv},
      primaryClass={cs.CL},
      url={https://arxiv.org/abs/2507.06261}, 
}

@inproceedings{DBLP:journals/corr/abs-2502-09211,
  author       = {Jakob Johannes Bauer and
                  Thomas Eiter and
                  Nelson Higuera Ruiz and
                  Johannes Oetsch},
  title        = {Visual Graph Question Answering with {ASP} and LLMs for Language Parsing},
  booktitle    = {{ICLP}},
  series       = {{EPTCS}},
  volume       = {416},
  pages        = {15--28},
  year         = {2024},
  month        = feb
}

@article{DBLP:journals/jair/VoborilRS25,
  author       = {Florentina Voboril and
                  Vaidyanathan Peruvemba Ramaswamy and
                  Stefan Szeider},
  title        = {Generating Streamlining Constraints with Large Language Models},
  journal      = {J. Artif. Intell. Res.},
  volume       = {84},
  year         = {2025}
}

@article{DBLP:journals/corr/abs-2512-17093,
  author       = {Timo Pierre Schrader and
                  Lukas Lange and
                  Tobias Kaminski and
                  Simon Razniewski and
                  Annemarie Friedrich},
  title        = {A Solver-in-the-Loop Framework for Improving LLMs on Answer Set Programming
                  for Logic Puzzle Solving},
  journal      = {CoRR},
  volume       = {abs/2512.17093},
  year         = {2025}
}

@article{DBLP:journals/ki/CalimeriGPRR18,
  author       = {Francesco Calimeri and
                  Stefano Germano and
                  Eliana Palermiti and
                  Kristian Reale and
                  Francesco Ricca},
  title        = {Developing {ASP} Programs with {ASPIDE} and LoIDE},
  journal      = {K{\"{u}}nstliche Intell.},
  volume       = {32},
  number       = {2-3},
  pages        = {185--186},
  year         = {2018},
  url          = {https://doi.org/10.1007/s13218-018-0534-z},
  doi          = {10.1007/S13218-018-0534-Z},
  bibsource    = {dblp computer science bibliography, https://dblp.org}
}
\clearpage

\end{document}